\documentclass[times]{elsarticle}
\usepackage{moreverb} 
\usepackage[colorlinks,bookmarksopen,bookmarksnumbered,citecolor=red,urlcolor=red,pdfauthor=author]{hyperref}
\usepackage{amsfonts,amsmath,amssymb}
\usepackage{psfrag,pstricks,pst-node}

\usepackage[section]{placeins}
\usepackage{subfigure} 
\usepackage{mathrsfs}
\usepackage{graphicx}
\usepackage[ruled,vlined]{algorithm2e}
\usepackage{fancyhdr}  
\usepackage{psfrag} 
\usepackage[export]{adjustbox} 
\usepackage{enumerate} 
\usepackage{comment}

\usepackage{algorithmic}
\usepackage{bm}
\usepackage{booktabs}
\usepackage{threeparttable}
\graphicspath{{./figs/}{./model structure/}}
\DeclareGraphicsExtensions{.pdf,.eps}

\usepackage{pgfplots}
\pgfplotsset{compat=1.13}

\begin{document}
 
\begin{frontmatter}
\renewcommand{\thefootnote}{\fnsymbol{footnotemark}}

\fancypagestyle{plain}{%
\fancyhf{} 
\fancyhead[RO,RE]{\thepage} 
}

\title{A data driven reduced order model of fluid flow by Auto-Encoder and self-attention deep learning methods}
    \author[lab1]{R. Fu}
    \author[lab1]{D. Xiao\corref{cor1}} 
    \cortext[cor1]{Corresponding author}
    \ead{dunhui.xiao@swansea.ac.uk} 
    \author[lab2]{I.M. Navon} 
     \author[lab1]{C. Wang}
    \address[lab1]{ZCCE, Faculty of Science and Engineering, Swansea University, Bay Campus,Swansea, UK, SA1 8EN} 
    \address[lab2]{Department of Scientific Computing, Florida State University, Tallahassee, FL, 32306-4120, USA}

\begin{abstract}
This paper presents a new data-driven non-intrusive reduced-order model(NIROM) that outperforms the traditional Proper orthogonal decomposition (POD) based reduced-order model. This is achieved by using Auto-Encoder(AE) and attention-based deep learning methods. The novelty of the present work lies in that it uses Stacked Auto-Encoder(SAE) network to project original high-dimensional dynamical systems onto a low dimensional nonlinear subspace and predict the fluid dynamics using an attention-based deep learning method. A new model reduction neural network architecture for fluid flow problem is presented. The SAE network compresses high-dimensional physical information into several much smaller sized representations in a latent space. These representations are expressed by a number of codes in the middle layer of SAE neural network. Then, those codes at different time levels are trained to construct a set of hyper-surfaces with multi variable response functions using attention-based deep learning methods. The inputs of the 
attention-based network are previous time levels’ codes and the outputs of the network are current time levels’ codes. The codes at current time level are then projected back to the original full space by the decoder layers in the SAE network. 

The capability of this data-driven reduced order model (DDROM) is illustrated numerically by two test cases: flow past a cylinder, and a lock exchange, respectively. The results obtained show that the DDROM performs better in terms of accuracy than the popular model reduction method namely proper orthogonal decomposition.
The improvement is assessed by comparison with a high fidelity POD model.

\end{abstract}

\end{frontmatter}
 
\section{Introduction}
\vspace{-2pt}
In engineering and physics, most physical phenomena are governed by conservation laws, which can be written as partial differential equations (PDEs) \cite{hijazi2020data}. However, solving PDEs can be computationally intensive in particular for high-fidelity realizations\cite{rudy2017data}. In this case, the reduced-order models(ROMs) technology plays an important role as they are able to simulate physical systems accurately with several orders of magnitude CPU speed-up.
The ROMs have been applied successfully to a number of research fields such as data assimilation \cite{vermeulen2006model}, ocean modelling \cite{cao2006reduced}, shallow water equations \cite{daescu2008dual,cstefuanescu2013pod}, air pollution prediction \cite{XIAO2019463}, polynomial systems \cite{benner2021interpolation}, viscous and inviscid flows \cite{martini2018certified}, large-scale time-dependent systems \cite{peherstorfer2017data}, optimal control \cite{alla2018posteriori},  circuit systems \cite{son2015model,hachtel2018multirate}, inverse problems \cite{cstefuanescu2016model}, fluids \cite{xie2018data}, reservoir history matching \cite{xiao2021efficient} and turbulent flows \cite{carlberg2013gnat}.

Among traditional model reduction methods, proper orthogonal decomposition (POD) combined with Galerkin projection is a popular model reduction method, and has been applied successfully to a number of fields. However, it lacks stability and it is highly inefficient for non-linear models\cite{FLM9529305,FLM9241738,Chaturantabut2010}.  Several stabilization methods such as calibration \cite{carlberg2011efficient}, Regularization\cite{willcox2003model}, Petrov-Galerkin \cite{xiao2013non} have been presented. In addition, an empirical interpolation Method (EIM)\cite{barrault2004empirical}, discrete empirical interpolation Method (DEIM) \cite{Chaturantabut2010,chaturantabut2011application,chaturantabut2012state}, a combination of Quadratic expansion and DEIM (residual DEIM method) \cite{xiao2014non}, Petrov-Galerkin method \cite{carlberg2011efficient} and a Gauss–Newton method with approximated tensors (GNAT) \cite{carlberg2013gnat} have been presented aimed at alleviating the nonlinear inefficiency of POD. The DEIM combined with POD method has been used to improve nonlinear term reduction efficiency, which means it improves ROM's CPU cost, but it does not improve the accuracy of the POD method. 

More recently, in order to avoid the above mentioned  issues, a data-driven non-intrusive reduced order modelling (NIROM) method was presented and has been applied to a number of research areas such as \cite{xiao2015non,peherstorfer2016data,rahman2019nonintrusive,kast2020non,mou2021data}.  Xiao et al. presented a number of NIROMs based on POD and interpolation methods and machine learning or deep learning methods \cite{xiao2019domain, xiao2015non}. Wang et al. presented a NIROM based on POD and neural network and applied it to a combustion problem \cite{wang2019non}. Jacquier et al. presented a NIROM based on deep neural network and POD and applied it to a flooding problem \cite{jacquier2021non}. Ahmed presented a NIROM for non-ergodic flows using POD and a long short term memory neural network together with a principal interval decomposition  \cite{ahmed2019memory}. 

As mentioned above, POD based model reduction methods restrict the state to evolve in a linear subspace (linear trial sub-spaces), which imposes a fundamental limitation on the efficiency  and accuracy of the resulting ROM \cite{lee2020model}. These linear trial sub-spaces also exist in other model reduction methods \cite{lee2020model}, such as balanced truncation \cite{kurschner2018balanced}, rational interpolation \cite{gugercin2008h_2} and reduced-basis method \cite{pagani2018numerical}.

In this paper, we present a new data driven non-intrusive model reduction framework using Stacked Auto-Encoder(SAE) network and attention-based deep learning methods to tackle the above linear trial sub-spaces issue. The new data-driven NIROM uses Stacked Auto-Encoder(SAE) network to project original high-dimensional dynamical systems into a nonlinear subspace and predict the fluid dynamics using an attention-based deep learning method. 

Deep learning method is a popular artificial neural network method as it has shown great potential in many research areas such as image recognition, materials, facial recognition, speech recognition, drug discovery, self-driving cars\cite{lecun2015deep,vamathevan2019applications,tagade2019attribute}. Auto-Encoder is a type of artificial neural network in which the input layer has the same dimensional size and data as the output layer. The high-dimensional inputs are passed into the network and are compressed in the network. The middle layer of the network has a smaller number of neurons compared to the input and output layers. Thus, the middle layer represents the inputs using reduced number of neurons (or referred to as 'codes' in Auto-Encoder network)\cite{tran2021fast}. Auto-Encoder has been applied to a number of areas such as image compression and denoising. Dimensionality reduction is the main application of Auto-Encoder network. Auto-Encoder has recently applied to ROM \cite{phillips2021autoencoder, wu2021reduced}. In the work of \cite{phillips2021autoencoder}, Auto-Encoder has been used for eigenvalue problems. In \cite{wu2021reduced}, the convolutional Auto-Encoder combined with self-attention has been used to reduce the dimensional size of the fluids images and Long short-term memory (LSTM) is used to predict the temporal fluid dynamics after reduction.

The other deep learning method used in this work is the attention mechanism. It was presented to deal with long sequence prediction problems. It imitates the process of organism observation behaviour, especially in humans, which means that after scanning the whole information or image, the target area is noticed with more attention while other areas are neglected to various degrees \cite{waswani2017attention}. In attention-based models, each item in the sequence has been assigned an attention score through the computing of global Query vector(Q) and Key vector(K). These scores are weighted summation with Value vector(V) to print out a new vector of the same shape. The new vector contains the long-term and local dependencies of items among the whole sequence. That makes the information processing based on these dependencies possible. The attention mechanism has been used to construct a transformer deep learning method and has been used in natural language processing and in computer vision \cite{phan2021self,zhao2020exploring}. 

In this work, we firstly use attention-based deep learning architecture to construct a multi-variable response function (hypersurface) to predict the fluid dynamics in a latent space. The new presented data-driven Reduced Order Model (DDROM) is implemented in the framework of FLUIDITY \cite{amcg_2015}. The Fluidity software is an open source, unstructured mesh, finite element computational fluid dynamics (CFD) three dimensional model and it is capable of numerically solving the Navier-Stokes (NS) equations. In order to illustrate the performance of this novel data-driven NIROM, two test cases, flow past a cylinder and a lock exchange are illustrated. In addition, the performance of this DDROM is compared against the solutions of proper orthogonal decomposition (POD) based ROM. The numerical results show that the new DDROM is capable of capturing the details of flows while the CPU time is reduced by several orders of magnitude. In addition, it performs better than the POD based ROM.

The structure of the paper is as follows. A detailed introduction of the governing equations, training system is shown in Section~\ref{Governing equations}. Section~\ref{nirom} provides a brief overview of SAE and self-attention, and their implementation in ROM for this specific problem. Section~\ref{examples} illustrates the performance of this ROM for two test cases: flow past a cylinder and lock exchange. Finally, in section~\ref{Conclusions}, the summary and conclusions are presented.

\section{Governing equations} \label{Governing equations}
The three dimensional (3D) non-hydrostatic and continuity NS equations which describe the conservation of momentum and mass of fluids are given by 
	\begin{eqnarray}
		\frac{\partial {\bm{u}}}{\partial t} + {\bm{u}}\cdot\nabla\bm{u} &=&
		-\nabla p  + \nabla\cdot \bm{\tau}, 
		\label{momeqn1} \\
		\quad \nabla\cdot\bm{u} &=& 0,
		\label{conteqn}
	\end{eqnarray}
	where $\bm{u}\equiv(u,v,w)^T$ is the velocity vector, $t$ is the time, $p=\tilde{p}/\rho_0$ is the normalised pressure ($\tilde{p}$ is the pressure and $\rho_0$ is the constant reference density), and $\bm{\tau}$ is the stress tensor.  
The discretised form of the system can be written as
    \begin{eqnarray}
        \label{momeqnc}
        \mathcal{C}^T \mathbf{u} = \mathbf{0} ,  \\
        \mathcal{M} \frac{\partial {\mathbf{u}}}{\partial t} + \mathcal{A}({\mathbf{u}}) {\mathbf{u}}  + \mathcal{K} {\mathbf{u}} + \mathcal{C}\mathbf{p} =  \mathbf{0},
    \end{eqnarray}\textbf{}
where $\mathcal{C}$ is a pressure gradient matrix, $\mathcal{M}$ denotes the mass matrix, $\mathcal{A}(\mathbf{u})$ denotes the solution dependent streaming operator and $\mathcal{K}$ denotes the matrix related to the remaining linear velocity terms. The velocity, $\mathbf{u}$, is a vector containing nodal values of all three components now, likewise, $\mathbf{p}$ is a vector containing the pressure nodal values. 

\section{Data-driven reduced order model (DDROM)}\label{nirom}
This section describes the construction and use of DDROM using Auto-Encoder neural network and attention-based deep learning method.

\subsection{Stacked Auto-Encoder neural network} \label{sec-sae}
\begin{figure}[h]
\centering 
\begin{minipage}{\linewidth}
\includegraphics[ width=\linewidth,angle=0,clip=true]{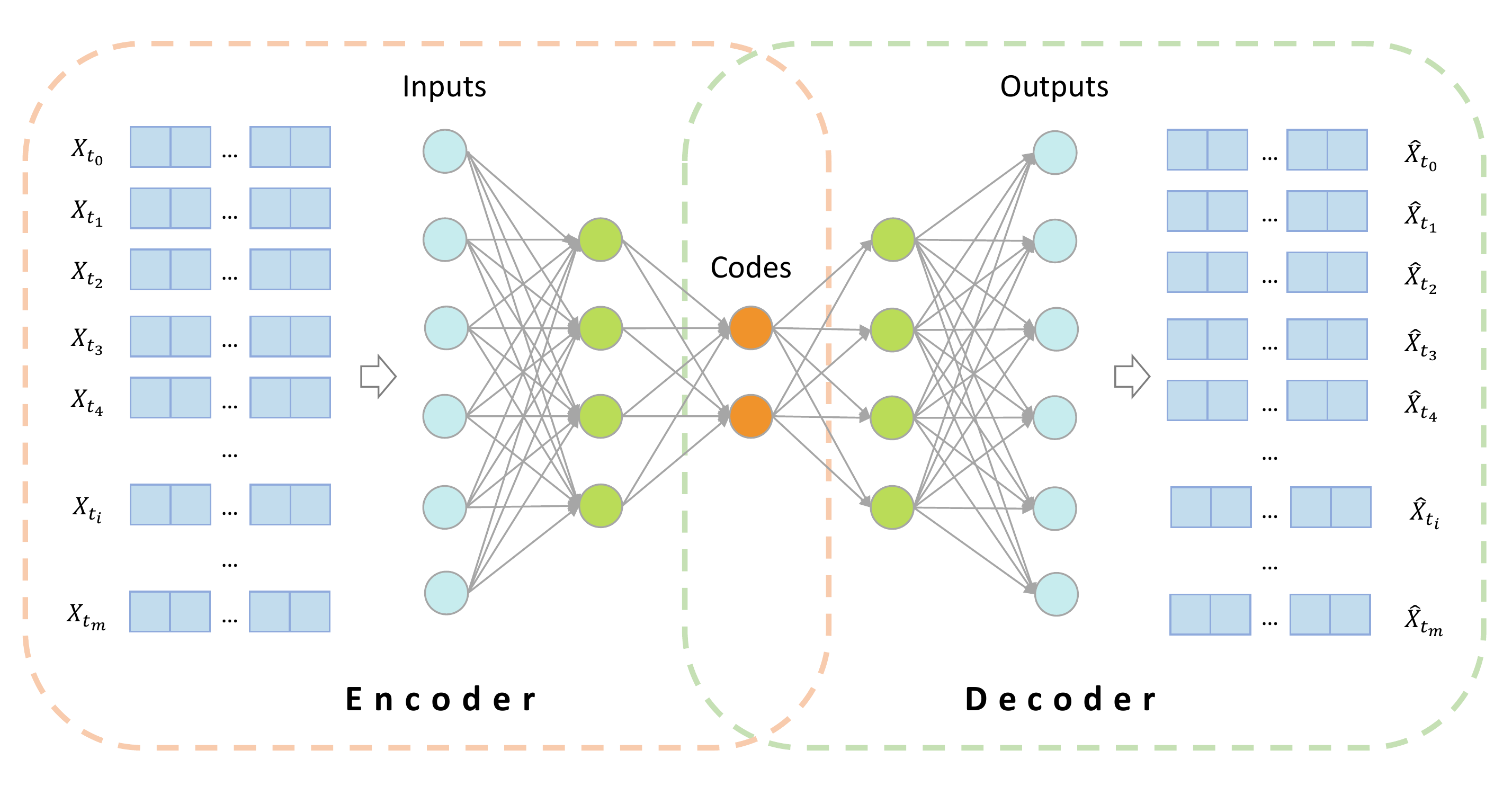} 
\end{minipage}
\caption{SAE Structure}
\label{SAE-str}
\end{figure}

The basic Auto-Encoder(AE) is an unsupervised feed-forward neural network that can be used to reduce the dimensional size.  Instead of labelling the outputs for training, the AE network sets the output the same number of nodes (neurons), shape and values as the inputs. It first maps the input data into a reduced dimensional latent space (represented by a number of codes in the middle layer of the neural network) and then projects the latent representation with reduced dimensional size to the output, see Figure \ref{SAE-str}. 

 The purpose of this particular neural network architecture is to reconstruct the input data into reduced dimensional size. The input data will be velocity $\mathbf{u}$,  pressure $\mathbf{p}$ in flow past a cylinder case and temperature for the lock exchange test case. An Auto-Encoder consists of two main parts, the encoder and the decoder, which can be described as transitions $\mathcal{E}$ and $\mathcal{D}$, such that:
\begin{equation}
\mathcal{E}:\mathbf{X} \rightarrow \mathcal{H}
\end{equation}
\begin{equation}
\mathcal{D}:\mathcal{H} \rightarrow \mathbf{X}
\end{equation}
\begin{equation}
\mathcal{E},\mathcal{D} = \underset{\mathcal{E},\mathcal{D}}{\operatorname{arg\,min}}\, \|\mathbf{X}-(\mathcal{E} \circ \mathcal{D}) \mathbf{X}\|^2
\end{equation}

In the simplest case (only one hidden layer), the encoder takes the input $\mathbf{x} \in  \{t_0,t_1,t_2,...,t_m\} = \mathbf{X}$ and maps it to $\bm{\alpha}  = \mathcal{H}$:
\begin{equation} \label{layw}
\bm{\alpha} = \sigma(\mathbf{Wx}+\mathbf{b}).
\end{equation}
In equation \ref{layw}, the $\bm{\alpha}$ is the code, or latent variables, or a reduced representation of the full system and will be used for training the self-attention deep neural network. These representations are similar to the POD coefficients in POD based ROM, but not identical.  
In the POD method, any variable $\mathbf{x}$ such as velocity or pressure can be stated by an expression: 
\begin{equation}
\mathbf{x} = \overline{\mathbf{x}} + \sum_{j=1}^{m} \alpha_j \bm{\phi}_j,
\end{equation}
where $\alpha_j$ represents the $j$th POD coefficient, $\bm{\phi}_j$ denotes the $j$th POD basis function and $\overline{\mathbf{x}}$ is the mean of snapshots for the variable $\mathbf{x}$. 
$\sigma$ in equation \ref{layw} is an activation function such as a 
 $sigmond$ function: 
 \begin{equation}
 \sigma (x)=1/(1+e^{-x}) 
 \end{equation}
 or $tanh$ function:
 \begin{equation}
 tanh(x)=(e^{x}-e^{-x})/(e^{x}+e^{-x}).
 \end{equation} 
 $\mathbf{W}$ in equation \ref{layw} is the weights vector and $\mathbf{b}$ is a bias vector. The weights and biases are both initialised randomly, and then updated iteratively using back-propagation.  Then, at the decoder stage,  the Auto-Encoder maps $\bm{\alpha}$ to $\mathbf{x'}$ with the same shape as the input $\mathbf{x}$:
\begin{equation} \label{aedecoder}
 \mathbf{x'} = \sigma'(\mathbf{W'} \bm{\alpha}+\mathbf{b'}), 
\end{equation}
where $\mathbf{W'}$, $\sigma'$ and $\mathbf{b'}$ are weights, activation function and bias vector respectively for the decoder. 
The Auto-Encoder is trained to minimise the errors between the inputs $\mathbf{x}$ and the reconstruction $\mathbf{x'}$. The loss function or cost function minimising the errors is then can be described as:
\begin{equation}
 \mathcal{L}(\mathbf{x},\mathbf{x'})=\|\mathbf{x}-\mathbf{x'}\|^2=\|\mathbf{x}-\sigma'(\mathbf{W'}(\sigma(\mathbf{Wx}+\mathbf{b}))+\mathbf{b'})\|^2 
\end{equation}
 
The stacked Auto-Encoder(SAE) is a neural network architecture with more then one hidden layers. 

\subsection{Attention-based deep learning method}\label{sec-attention} 
Self-attention was proposed by Vaswani et al. to deal with the long sequence prediction challenges faced by the encoder-decoder network \cite{waswani2017attention}. The self-attention mechanism allows the inputs to interact with each other ('self' process) and find out which should be paid more attention ('attention' process) and it shows great potential in long sequence prediction.  
In our work, we use the self-attention mechanism and weights each part of the input data differently. Unlike previous popular sequential models, the self-attention mechanism does not process the input data in order. For example, the code values $\bm{\alpha}_0,\bm{\alpha}_2,\bm{\alpha}_2,...,\bm{\alpha}_{\tau}$ generated by Auto-Encoder network at different time levels: $t=0,t=1,..., t=\tau$ are not necessarily inputted into the network in order. $\tau$ is the input number of time levels for a training data pair(input-output). It gives different weights to each time level of the code values $\bm{\alpha}_i$ ($i$ is an arbitrary time level).

The attention mechanism can be described by the following equation:
\begin{equation}
    Attention(Q,K,V)=softmax(\frac{QK^{\mathsf{T}}}{\sqrt{d_{k}}})V
\end{equation}
 Q denotes a query matrix, which is essentially a vector representation of one snapshot at a time level in the input sequence. K is a key matrix, which is a vector representations of all of the snapshots in the input sequence.  V denote the values, which are again the vector representations of all of the snapshots in the input sequence. $d_{k}$ is a scaling factor.
Since the values in V are multiplied and summed with some attention-weights w, it can be simplified as, 
 \begin{equation}
    w = softmax(\frac{QK^{\mathsf{T}}}{\sqrt{d_{k}}})
\end{equation}
 This means that the weights w are defined by how each snapshot of the sequence (represented by Q) is influenced by all the other snapshots in the sequence (represented by K).

\begin{figure}[h]
\centering 
\begin{minipage}{0.8 \linewidth}
\includegraphics[width = \linewidth,angle=0,clip=true]{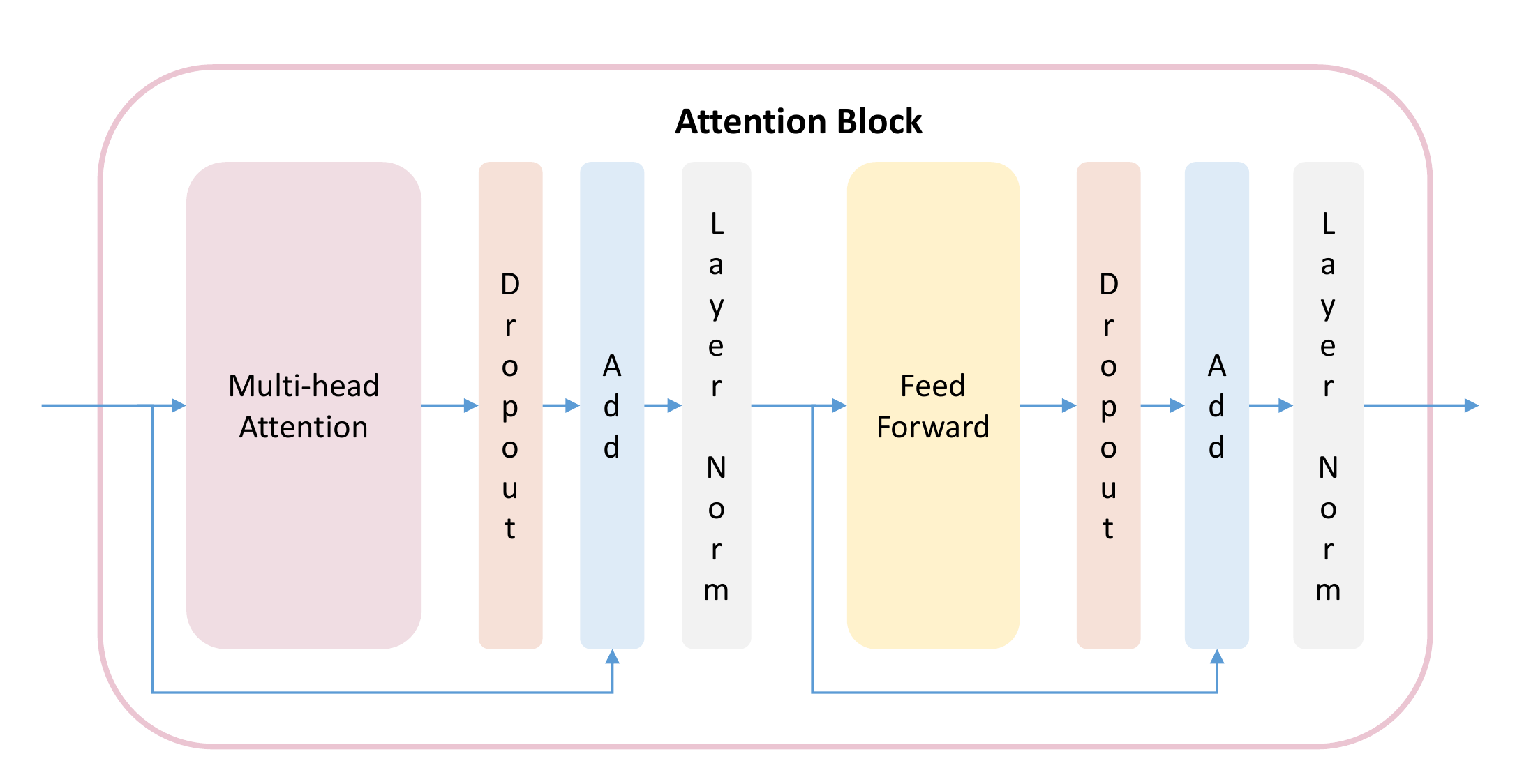} 
\end{minipage}
\caption{The figure shows attention block in the ROM architecture}
\label{tf_encoder}
\end{figure}

\begin{figure}[h]
\centering 
\begin{tabular}{cc}
\begin{minipage}{0.7\linewidth}
\includegraphics[width = \linewidth,angle=0,clip=true]{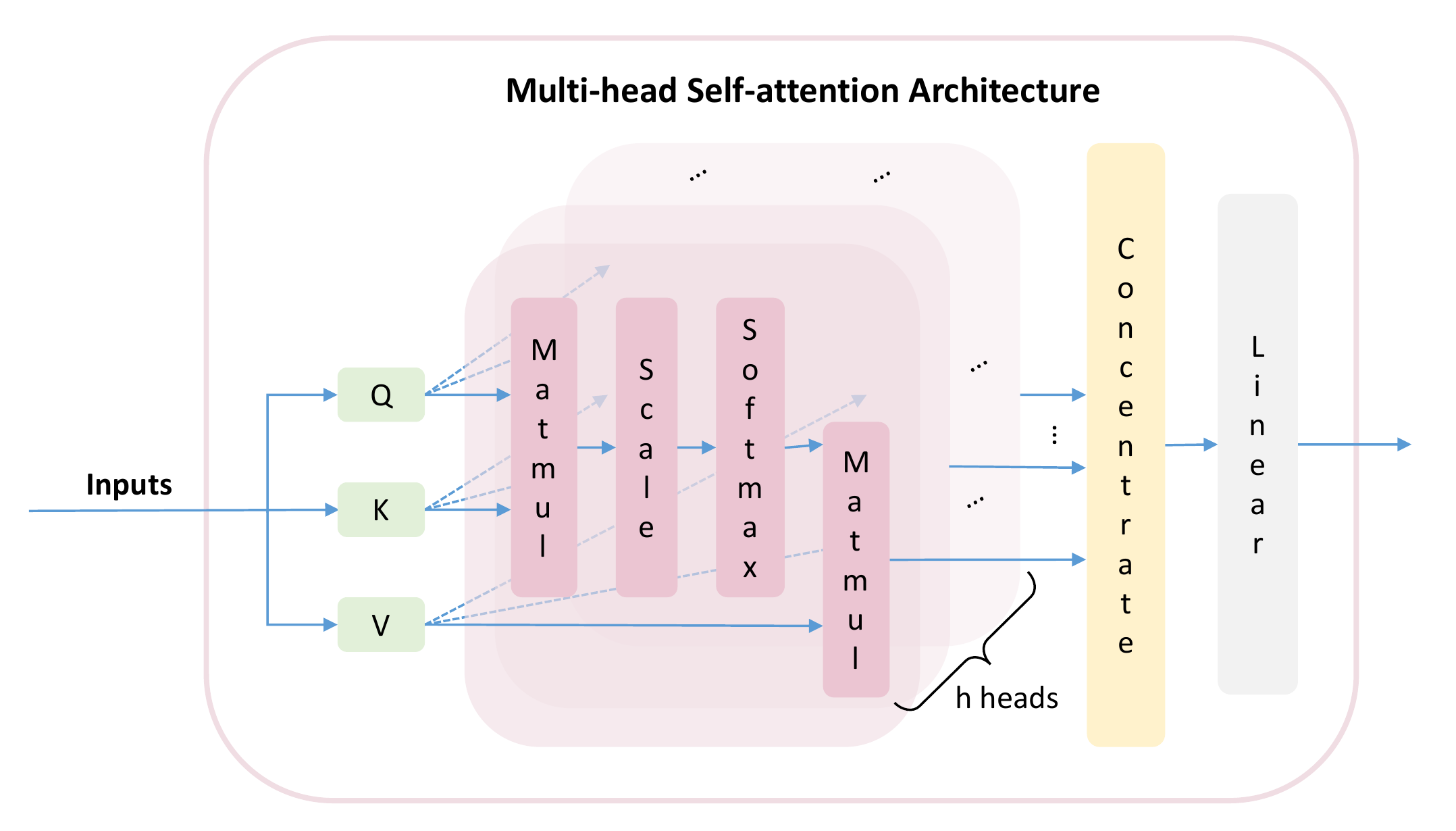}  
\end{minipage}
&
\begin{minipage}{0.25\linewidth}
\includegraphics[width = \linewidth,angle=0,clip=true]{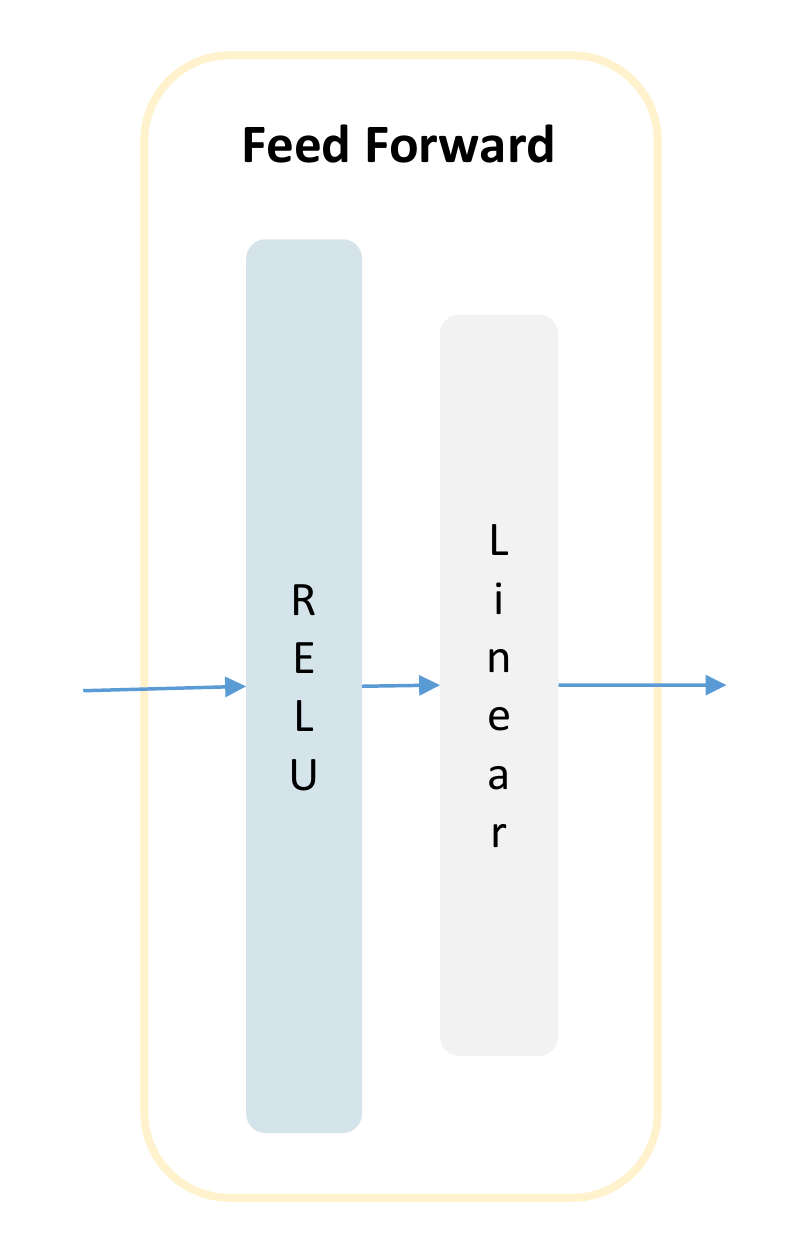}
\end{minipage}\\ 
(a) {\small Multi-head attention architecture}&
(b) {\small Feed Forward Block}\\
\label{multi-ff}
\end{tabular}
\caption{Multi-head attention architecture and Feed forward block architecture}
\end{figure}

In this work, the idea of self-attention mechanics is used and the architecture that represents fluid dynamics for our reduced order model is given in the Figure \ref{tf_encoder}.  In this architecture, six encoder blocks are main parts and it is similar to only encoder part of a complete transformer architecture, which is a sequence prediction. The attention block consists of two main parts: multi-head attention (Figure \ref{multi-ff} (a)) and feed forward layers (Figure \ref{multi-ff} (b)).  

\subsection{Data-driven reduced order modelling based on Auto-Encoder and Self-attention}\label{dldrrom}
\begin{figure}[h]
\centering 
\includegraphics[width = \linewidth,angle=0,clip=true]{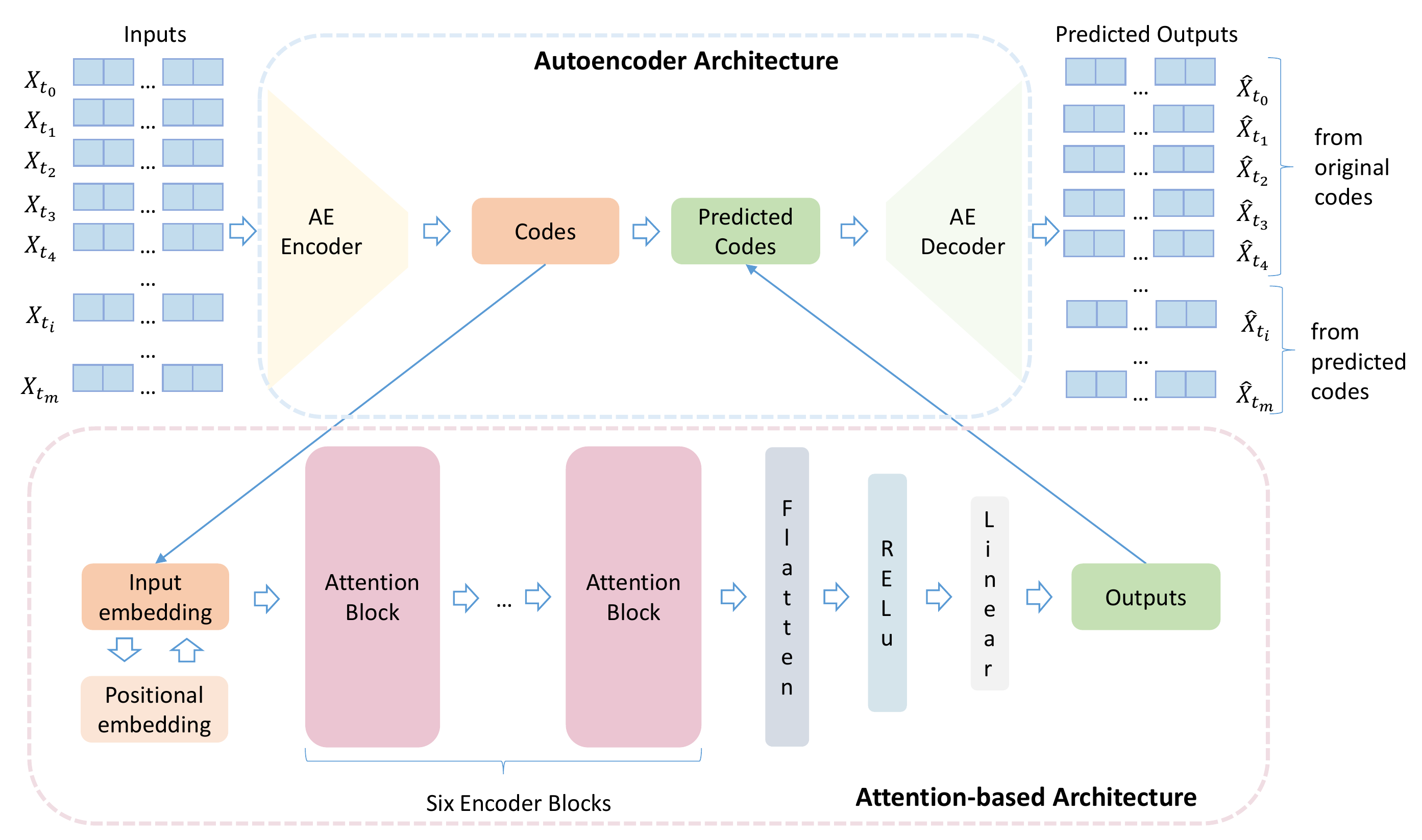} 
\caption{The figure shows neural network architecture for DDROM}
\label{mainarchi}
\end{figure}

This work presents a novel neural network architecture for reduced order model. The architecture can be described in Figure \ref{mainarchi}. This new data driven reduced order model neural network architecture consists of two main parts: Auto-Encoder network and self-attention part. The Auto-Encoder network (above part in the figure) is constructed for projecting the full high-dimensional space into a reduced space (latent space). The multi-head scaled-dot self-attention part (bottom part in the figure) is a structure that is used to represent the fluid dynamics in the reduced space. Its inputs are the codes from the middle layer of Auto-Encoder network and its outputs will be next time level's predicted codes and then they are projected back to the full space via decoder. It includes input embedding, positional embedding, attention blocks, flatten, Rectified Linear Unit (ReLU) and linear modules. This is similar to half of the complete transformer architecture (Encorder part of the transformer architecture).

\begin{figure}[htbp!]
\centering
\includegraphics[width = 0.8\linewidth,angle=0,clip=true]{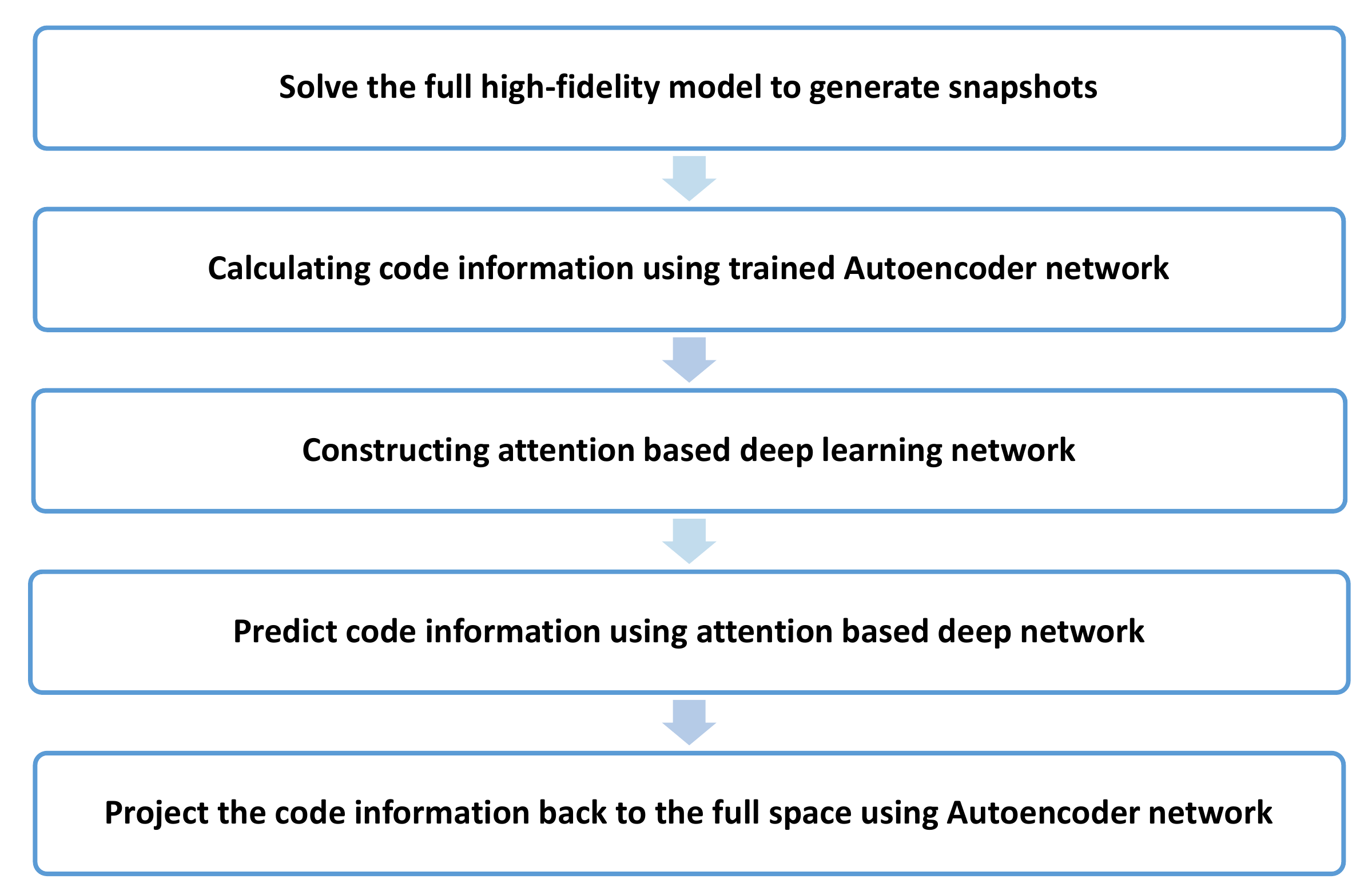} 
\caption{The main flowchart of DDROM}
\label{flow-chart-dd}
\end{figure}

This section describes the method of constructing a data-driven reduced order model (DDROM) using the Auto-Encoder and self-attention deep learning methods (offline stage) and how the DDROM predicts the fluid dynamics (online stage). The procedure can be summarized in the flowchart given in Figure \ref{flow-chart-dd}.
 
\subsection{Constructing the DDROM (offline stage)}  \label{constructing-offline}
 
Having obtained the codes in the SAE network (middle layer of the SAE network) in section~\ref{sec-sae}, the offline stage of DDROM will be completed by approximating the governing equations. This is achieved by using trained attention-based deep neural network to predict how the governing equations would behave. All simulation results generated by Fluidity at different time levels during a time period are projected onto the reduced space by the Auto-Encoder neural network architecture. The dimensional size of the codes in the middle layer of SAE is much smaller than that of the inputs, thus decreasing the dimensional size of the problem drastically. The codes information is similar to the POD coefficients in the POD method. The weights of neurons in other layers (excluding input, output and middle layers) store the projection information (projecting full original space into a reduced space and projecting back), which are similar to the POD basis functions in the POD method, see \cite{xiao2015nonrbf}. The codes are then used to train the self-attention deep neural network. This training results in a function~$f_j$ for each code value (reduced coefficient), which maps the set of code values from $\tau$ time levels ($\bm{\alpha}^{k-\tau-1}, \bm{\alpha}^{k-\tau},\ldots,\bm{\alpha}^{k-1}$) to the associated code value at the next time level ($\alpha_j^{k}$), i.e.
\begin{eqnarray}
\alpha_j^{k} = f_j(\bm{\alpha}^{k-\tau-1}, \bm{\alpha}^{k-\tau},\ldots,\bm{\alpha}^{k-1}) \nonumber\\
= f_j(\alpha_1^{k-\tau-1}, \alpha_2^{k-\tau-1}, \ldots, \alpha_m^{k-\tau-1}, \alpha_1^{k-\tau}, \alpha_2^{k-\tau}, \ldots, \alpha_m^{k-\tau},\ldots, \alpha_1^{k-1}, \alpha_2^{k-1}, \ldots, \alpha_m^{k-1}), \\
\qquad \forall k \in \{1,2, \ldots, \mathcal{N}_s \}\,.\nonumber
\label{reduced-generalrom}
\end{eqnarray}
where $\tau$ denotes input data length in training data. $\alpha_j$ denotes the $j^{th}$ code information for SAE method. $\alpha_j$ can be POD coefficients if POD method is used. $m$ denotes the number of codes used in the SAE network and it means the dimensional size. $\mathcal{N}_s$ denotes the total number of time levels, and it equals to the total number of snapshots. 
By including the initial condition, we have $\mathcal{N}_s$ pairs of input and output data that are used to form the response function~$f_j$ 
\begin{eqnarray}
\textrm{input:}&& \bm{\alpha}^{k-\tau-1}, \bm{\alpha}^{k-\tau},\bm{\alpha}^{k-\tau+1},\ldots,\bm{\alpha}^{k-1} \label{eq:GPRinputs} \\
\textrm{output:} && \alpha_j^{k}\,, \label{eq:GPRoutputs} 
\end{eqnarray}
for all $k\in\{1,2,\ldots,\mathcal{N}_s\}$. The bold $\bm{\alpha}$ is one complete code information vector ($\bm{\alpha}={\alpha_1,\alpha_2,...,\alpha_m}$). The input consists of $\tau$ time levels' complete code information. The output $\alpha_j^{k}$ denotes $j^{th}$ code information for time level $k$.  
This procedure is repeated for each code information (i.e.~for $j\in\{1,2,\ldots,m\}$), and once all the functions 
$\{f_j\}_{j=1}^{m}$ have been determined, the off-line stage is complete. 

\subsection{Running simulations with the data-driven reduced order model (online stage)} \label{online-simulation}
For running the DDROM, the functions $\{f_j\}_{j=1}^{m}$ are treated as response functions allowing the code information at one time level to be predicted given those at $\tau$ previous time levels

\begin{equation}
\alpha_j(t+\Delta t)= f_j(\bm{\alpha}(t),\bm{\alpha}(t-1),...,\bm{\alpha}(t-\tau+1)) \quad \forall j\in\{1,2,\ldots,m\}\,.
\end{equation}
We remark that when running the DDROM, the time step, $\Delta t$, will coincide with that of the high-fidelity full model. The procedure of on-line prediction using the DDROM is summarized in Algorithm~\ref{dd-online}.  The initial condition can be different than that used in the high-fidelity full model. The number of time steps ($\mathcal{T}$) can be different than that used in the training period. That is, the DDROM can be run for a longer or shorter time than the high-fidelity full model.\\
 
\begin{algorithm}[H] \label{dd-online}
\caption{On-line DDROM calculation}
\SetAlgoLined 
\vspace{3mm}
 {The response functions, $\left \{ f_{j} \right \}_{j=1}^{m}$ are known and Auto-Encoder network is already trained.} 
\\[5mm]
 {The initial condition ($\bm{\alpha}^0$), time step ($\Delta t$), initial time ($t_0$) and 
number of time steps ($\mathcal{T}$) are given.} 
\\[5mm]
\SetKwFor{FOR}{for}{do}{endfor} 
\FOR{$n = 1$ to $\mathcal{T}$}{
    $t = t_0+n\Delta t$ \ \ \  {current time} \\[5mm]
    {Step (a): calculate the code values, ${\bm{\alpha}}^n$, at the current time step:}\\
    \FOR{$j = 1$ to $m$}{ 
     
	$\alpha_j^n = f_j(\bm{\alpha}^{n-\tau-1}, \bm{\alpha}^{n-\tau},\ldots,\bm{\alpha}^{n-1} )$
 
  }
  \vspace{5mm}
   {Step (b): obtain the solutions, velocity $\mathbf{u}^n$ for example, in the full space at the current time, $t$, by projecting ${\alpha}_{j}^n$ back onto the full space using the Auto-Encoder network using Equation \ref{aedecoder}.}\\ 
  \vspace{5mm}
  
   $\mathbf{u}^n =  \sigma'(\mathbf{W'} \bm{\alpha}+\mathbf{b'})$
   \vspace{5mm}
 }
\end{algorithm}

\section{Illustrative numerical examples}\label{examples}
In this section, we demonstrate the capability of DDROM using two test problems, namely, a lock exchange and a 2D flow past a circular cylinder. The full model with simulation solutions of these two problems are generated via the finite element fluid model Fluidity\cite{pain2005three}. 
In both test cases, unstructured triangular meshes were used with sufficient resolution to ensure accurate solutions. Using this snapshot data the DDROM were constructed and then used to predict the problems.   

In this demonstration a comparison between Auto-Encoder based DDROM and POD based ROM has been carried out. In addition to comparing solution profiles, solution errors (root-mean-square errors(RMSE)) and correlation coefficient are analyzed. The formulation of CC can be described as,
 
\begin{equation}
CC(X^i(t),\hat{X}^i(t))=\frac{cov(X^i(t),\hat{X}^i(t))}{\sigma _{X^i(t)}\sigma _{\hat{X}^i(t)}}=\frac{E[(X^i(t)-\mu _{X^i(t)})(\hat{X}^i(t)-\mu _{\hat{X}^i(t)})]}{\sigma _{X^i(t)}\sigma _{\hat{X}^i(t)}},
\end{equation}
where $\hat{x}^{i}(t)$ and $x^{i}(t)$ are DDROM and high-fidelity full model
$i^{th}$ node solutions at time level $t$ respectively. The $\mu_{X^i(t)}$ and
$\mu _{\hat X^i(t)}$ are the expected values of $X^i(t)$ and $\hat{X}^i(t)$,
$\sigma _{X^i(t)}$ and $\sigma _{\hat{X}^i(t)}$ are standard deviations. The
RMSE measures the differences between solutions and it is computed as 
\begin{equation}
RMSE(t) = \sqrt{(\frac{1}{F})\sum_{i=1}^{F}(\hat{x}^{i}(t) - x^{i}(t))^{2}},
\end{equation}
where $F$ is the number of nodes in the computational domain.

\subsection{Case 1: Lock exchange}
\begin{figure}
\centering
\begin{minipage}{1\linewidth}
\includegraphics[width = \linewidth,angle=0,clip=true]{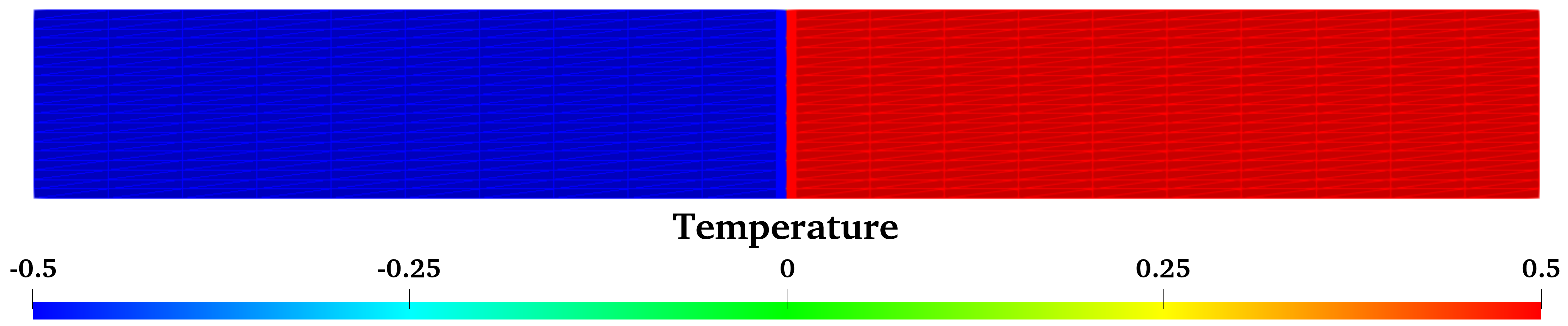} \end{minipage} \\
\caption{Case 1: Computational domain of a lock exchange case}
\label{le_mesh}
\end{figure}

In the first numerical example, a two dimensional lock exchange problem is demonstrated. The problem consists of cold and hot fluids with different densities, which are separated by a lock. The cold fluid is at the left and the hot fluid is at the right. Two gravity currents propagate along the tank after the lock is removed \cite{shin2004gravity}. The computational domain is a non-dimensional rectangle with a size of 0.8 $\times$ 0.1, see Figure \ref{le_mesh}. The mesh has 1490 nodes. The initial non-dimensional temperatures are set to be $T = - 0.5$ for the cold fluid and $T = 0.5$ for the hot fluid. The initial conditions for the pressure and velocity are set to be zero. The isotropic viscosity is $ 1 \times 10^{-10}$. The Crank–Nicolson method is used in the temporal discretization. In this work, temperature field evolution is simulated via the Fluidity. From the full model simulation by Fluidity, 10000 snapshots were obtained at regularly spaced time intervals $\Delta t$ = 0.01 for solution variables. The simulation period is 100s and $70\%$ of the simulation data is used for training the attention based deep neural network in order to predict the fluid dynamics. $10\%$ of the simulation data is used to validate the model. $20\%$ of the simulation data is used to test the model. The input number of time levels ($\tau$ in Equation \ref{reduced-generalrom}) is 24 in this case. 

\begin{figure}[htbp!]
\centering
\begin{adjustbox}{width={\textwidth},totalheight={\textheight},keepaspectratio}%
\begin{tabular}{cc}
\subfigure[Full Model, t = 10s]{
\begin{minipage}[t]{0.55\linewidth}
\centering
\includegraphics[width = \linewidth]{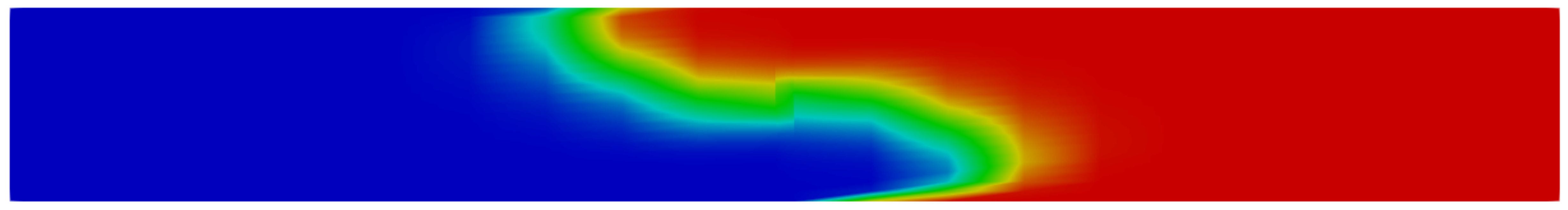}
\end{minipage}%
}%
\subfigure[Full Model, t = 45s]{
\begin{minipage}[t]{0.55\linewidth}
\centering
\includegraphics[width = \linewidth]{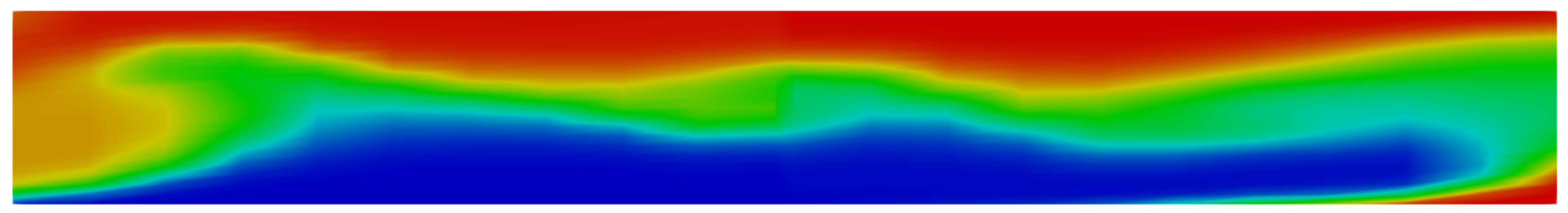}
\end{minipage}%
}%
\\
 
\subfigure[DDROM, code number = 5]{
\begin{minipage}[t]{0.55\linewidth}
\centering
\includegraphics[width = \linewidth]{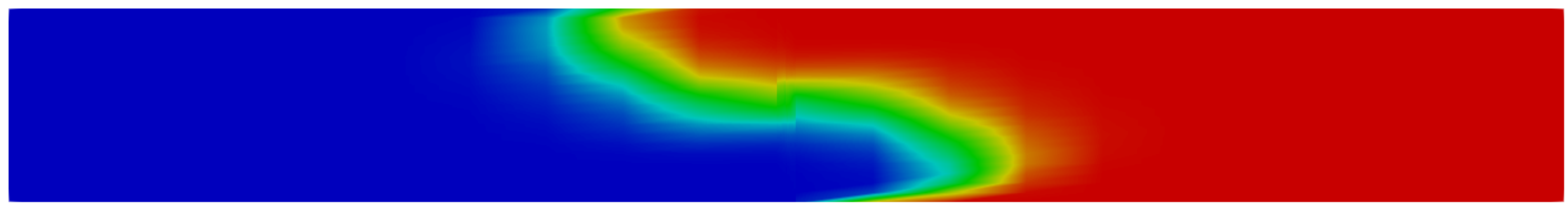}
\end{minipage}%
}%
\subfigure[DDROM, code number = 5]{
\begin{minipage}[t]{0.55\linewidth}
\centering
\includegraphics[width = \linewidth]{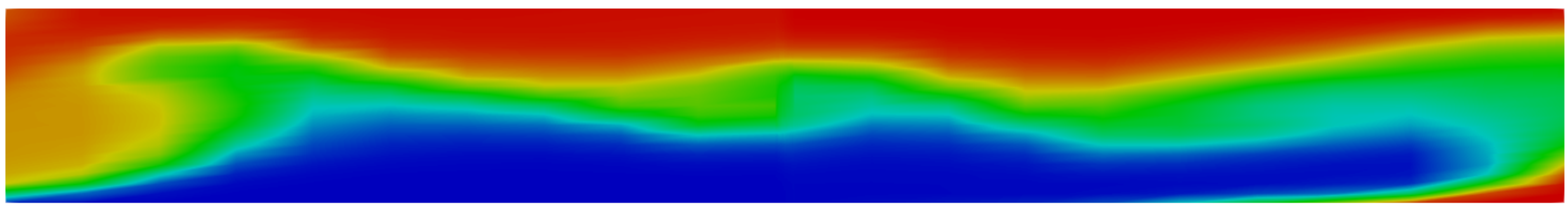}
\end{minipage}%
}%
\\
\subfigure[DDROM, code number = 8]{
\begin{minipage}[t]{0.55\linewidth}
\centering
\includegraphics[width = \linewidth]{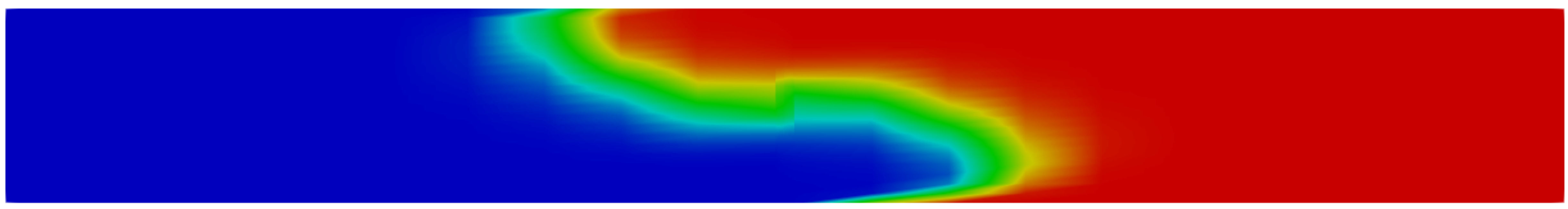}
\end{minipage}%
}%
\subfigure[DDROM, code number = 8]{
\begin{minipage}[t]{0.55\linewidth}
\centering
\includegraphics[width = \linewidth]{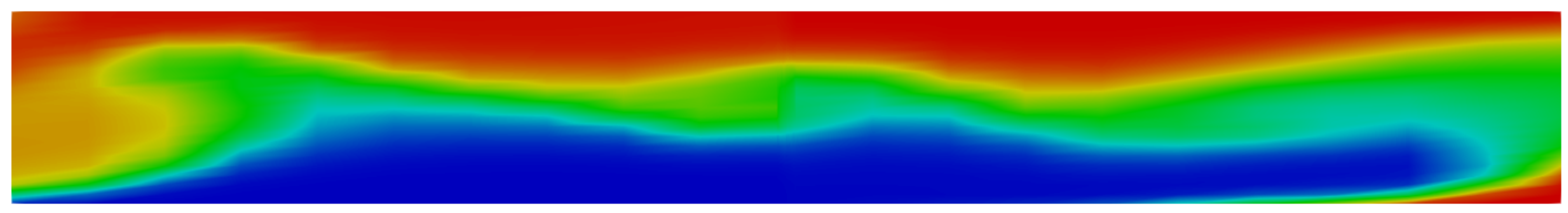}
\end{minipage}%
}%
\\
\subfigure[ROM with 8 POD basis functions]{
\begin{minipage}[t]{0.55\linewidth}
\centering
\includegraphics[width = \linewidth]{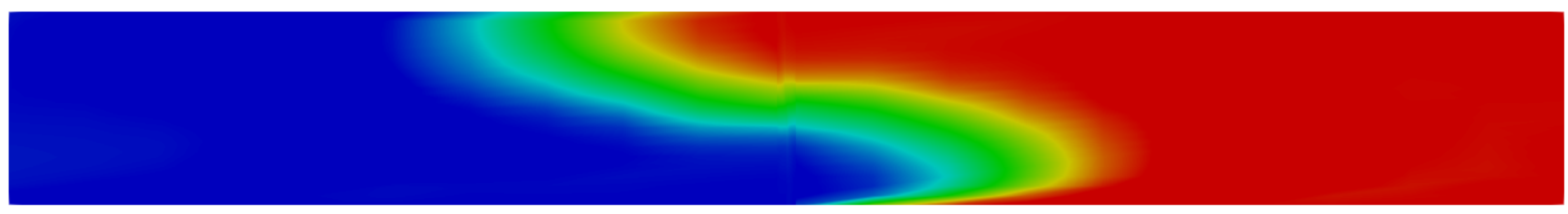}
\end{minipage}%
}%
\subfigure[ROM with 8 POD basis functions]{
\begin{minipage}[t]{0.55\linewidth}
\centering
\includegraphics[width = \linewidth]{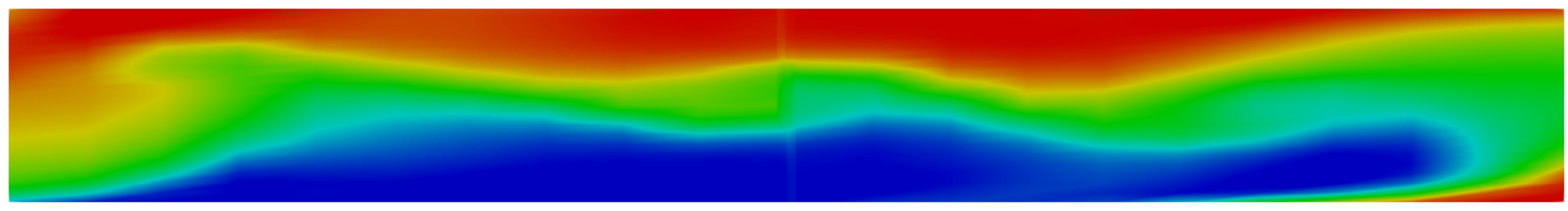}
\end{minipage}%
}%

\\

\begin{minipage}[t]{0.55\linewidth}
\centering
\includegraphics[width = \linewidth]{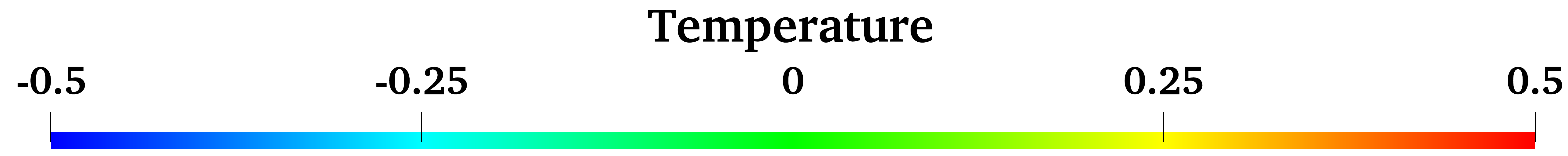}
\end{minipage}%

\begin{minipage}[t]{0.55\linewidth}
\centering
\includegraphics[width = \linewidth]{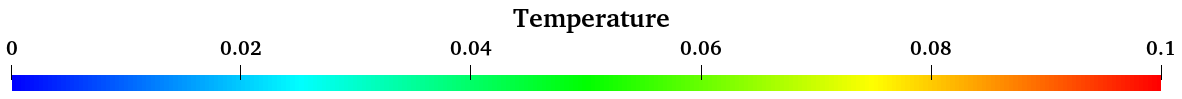}
\end{minipage}%

\\ 
\end{tabular}
\end{adjustbox}
\caption{Case 1: Lock Exchange. The temperature solutions obtained at time $t=10$s and $45$s from the full model, DDROM with 5 and 8 codes and ROM with 8 POD basis functions respectively.}
\label{lock-profile}

\end{figure}

Figure \ref{lock-profile} shows the temperature solutions obtained from the full model, DDROM with 5 and 8 codes and ROM with 8 POD basis functions at time $t=10$s and $45$s respectively. As shown in the figure, both of the Auto-Encoder based DDROM and POD based ROM can capture dominant structural details of the numerical solutions.  

\begin{figure}
\centering

\begin{minipage}{0.49\linewidth}
\includegraphics[width = \linewidth,angle=0,clip=true]{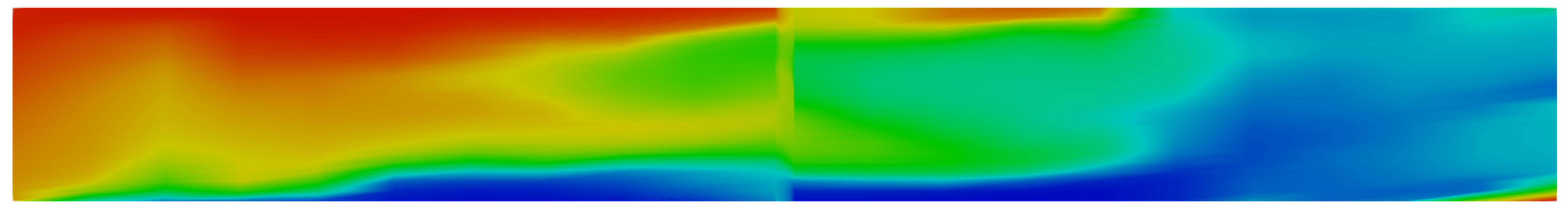}  
\end{minipage}\\ 
(a) {\small full model, t=85s }\\

\begin{minipage}{0.49\linewidth}
\includegraphics[width = \linewidth,angle=0,clip=rue]{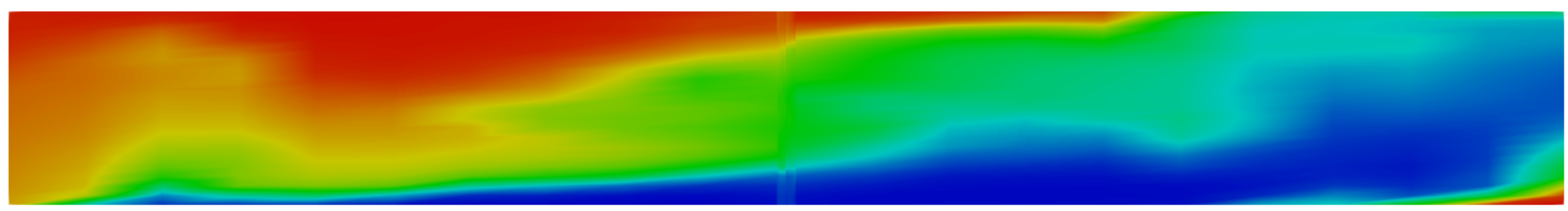}  
\end{minipage}\\ 
(b) {\small DDROM, code number = 5}\\

\begin{minipage}{0.49\linewidth} 
\includegraphics[width = \linewidth,angle=0,clip=true]{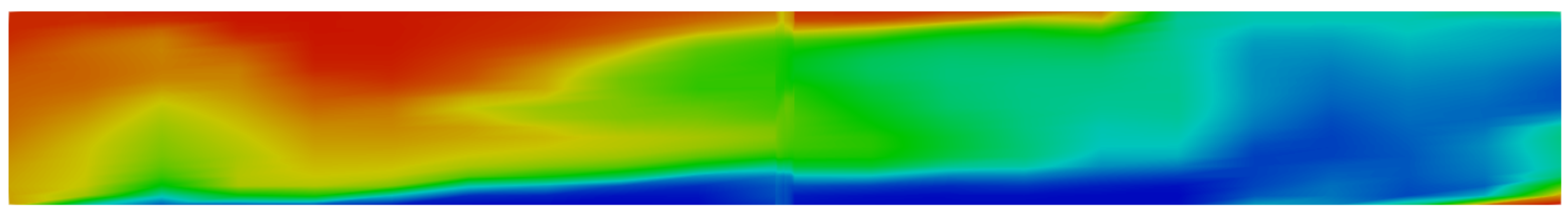}
\end{minipage}\\
(c) {\small DDROM, code number = 8}\\

\begin{minipage}{0.49\linewidth} 
\includegraphics[width = \linewidth,angle=0,clip=true]{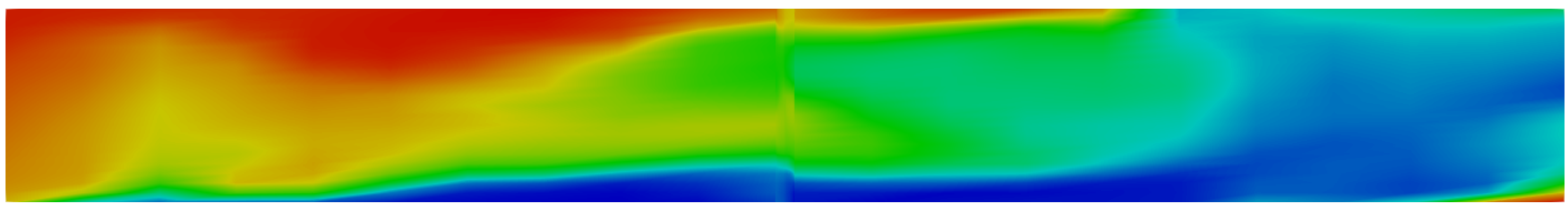}
\end{minipage}

(d) {\small ROM with 8 POD basis functions}\\  
\begin{minipage}[t]{0.49\linewidth}
\centering
\includegraphics[width = \linewidth]{le_legend}
\end{minipage}%

\caption{Case 1: lock exchange. The graphs (a)-(d) show the temperature solutions obtained from the full model, DDROM with 5 and 8 codes and ROM with 8 POD basis functions predicted time level $t=85s$.} 
\label{lock-predict85}
\end{figure}

\begin{figure}
\centering
\begin{minipage}{1\linewidth}
\includegraphics[width = \linewidth,angle=0,clip=true]{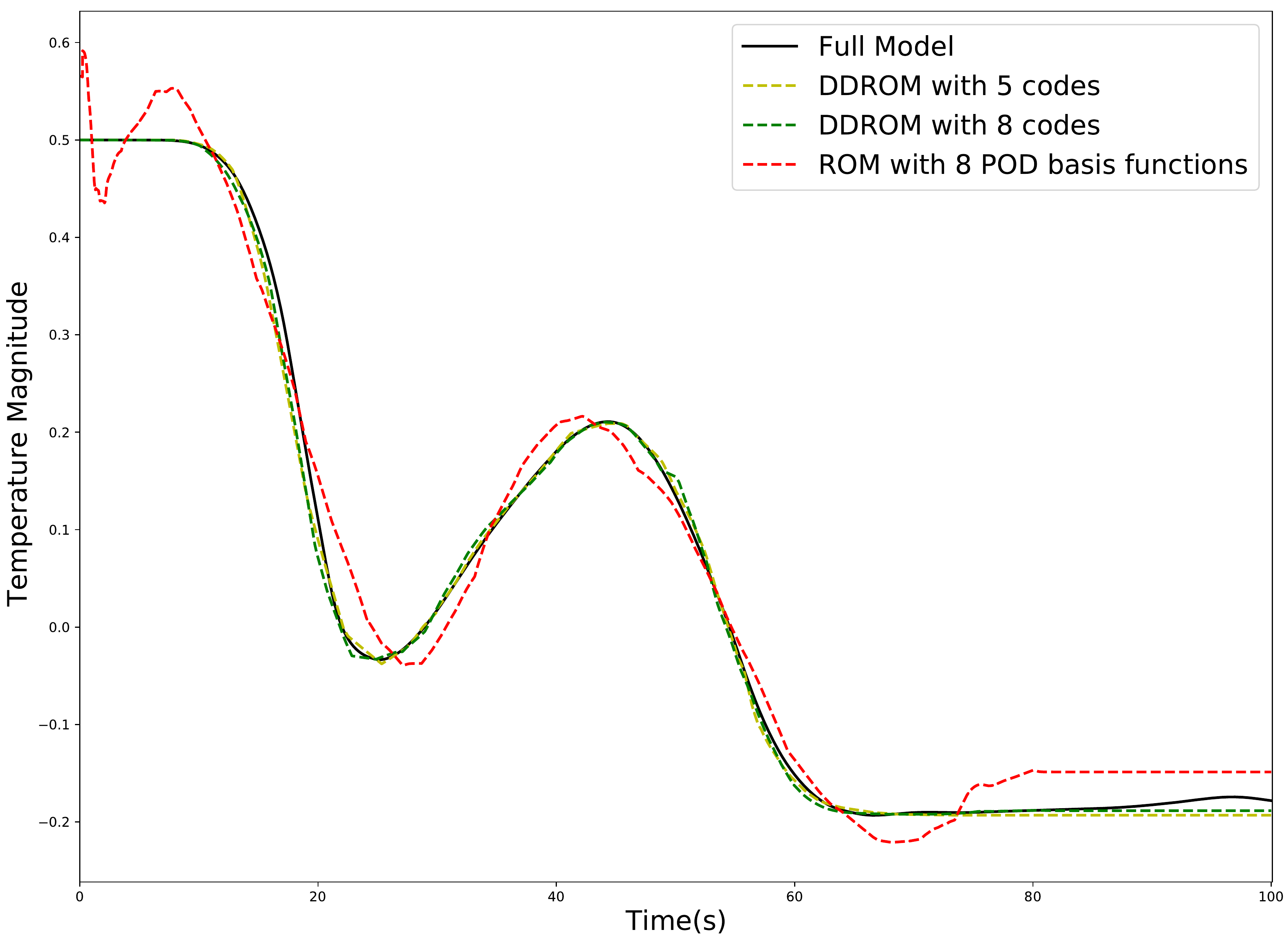} \end{minipage} \\
\caption{Case 1: lock exchange. The temperature solutions obtained from the full model, DDROM with 5 and 8 codes and ROM with 8 POD basis functions at a particular point $(x,y)=(0.4038, 0.095)$ in the computational domain.}
\label{le_point}
\end{figure}

\begin{figure}
\centering
\begin{tabular}{cc}
\begin{minipage}{0.55\linewidth}
\includegraphics[width = \linewidth,angle=0,clip=true]{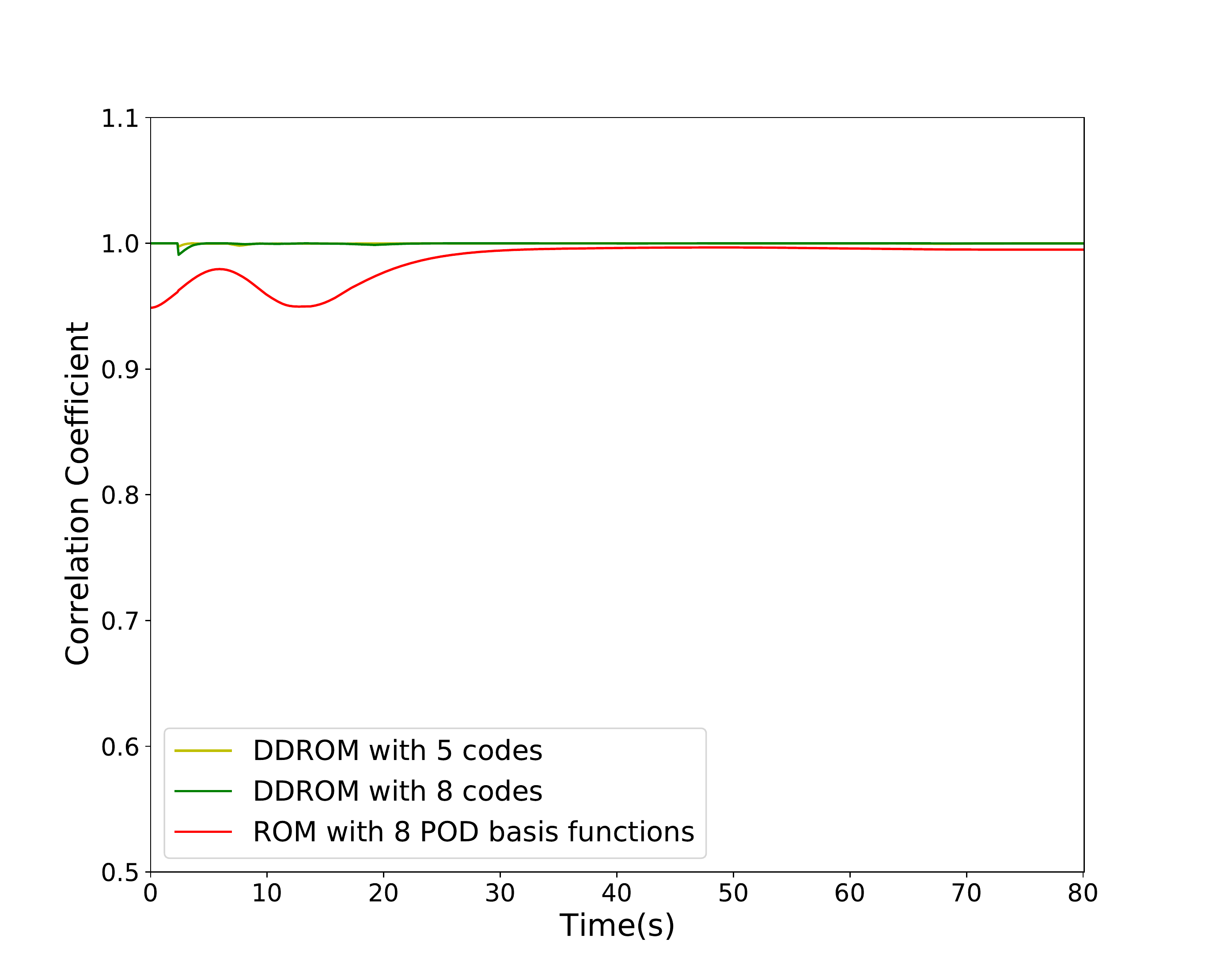}  
\end{minipage}
&
\begin{minipage}{0.49\linewidth}
\includegraphics[width = \linewidth,angle=0,clip=true]{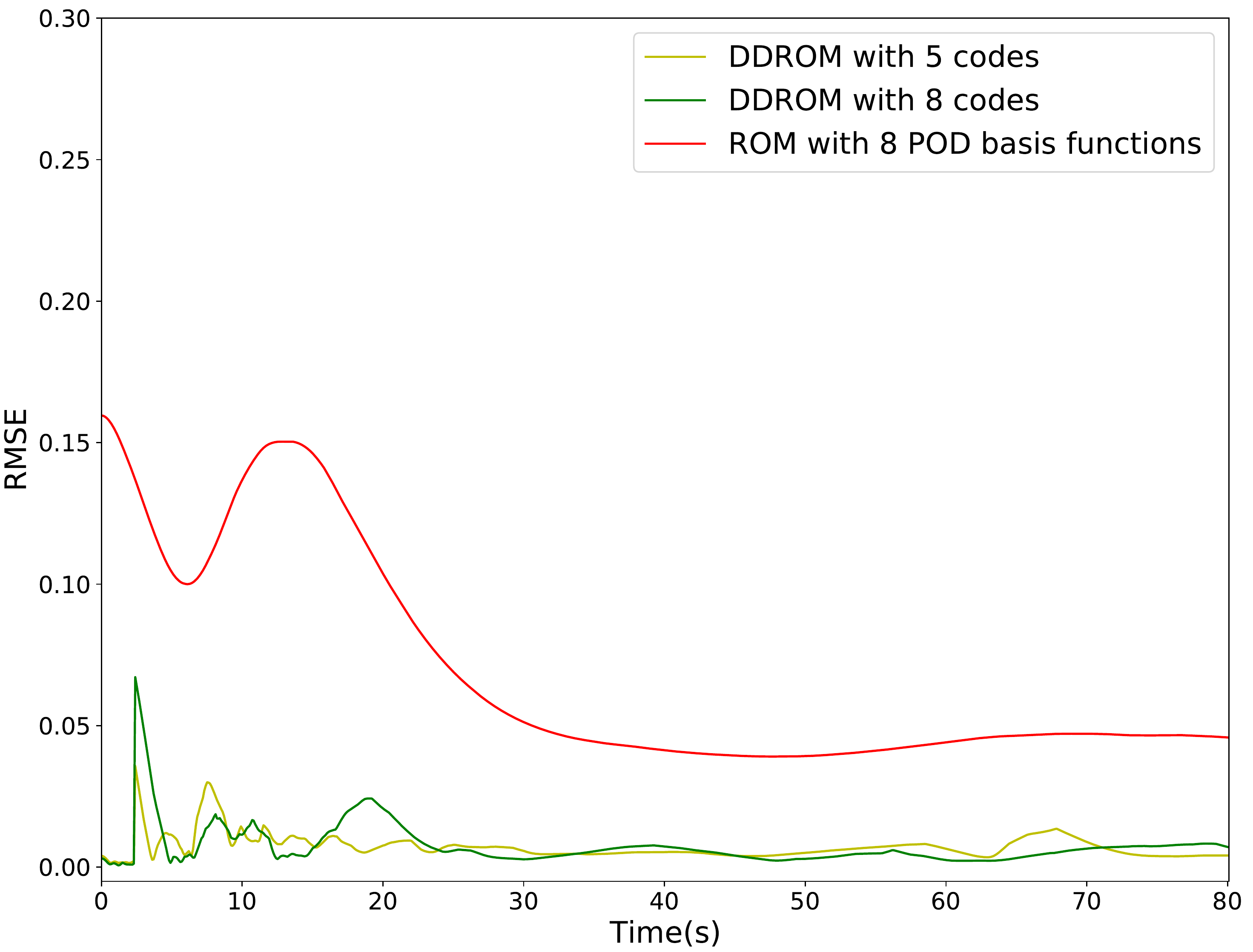}  
\end{minipage} \\
(a) {\small} Correlation Coefficient&
(b) {\small} RMSE\\
\end{tabular}
\caption{Case 1: Lock exchange. Correlation coefficient and the root-mean-square errors(RMSE) of temperature solutions from Auto-Encoder based DDROM with 5 and 8 codes and POD based ROM with 8 basis functions.}
\label{lock-cc-rmse}
\end{figure}

\begin{figure}[htbp!]
\centering
\begin{adjustbox}{width={\textwidth},totalheight={\textheight},keepaspectratio}%
\begin{tabular}{cc}
\subfigure[error from ROM with 8 POD basis functions, t = 40s]{
\begin{minipage}[t]{0.55\linewidth}
\centering
\includegraphics[width = \linewidth]{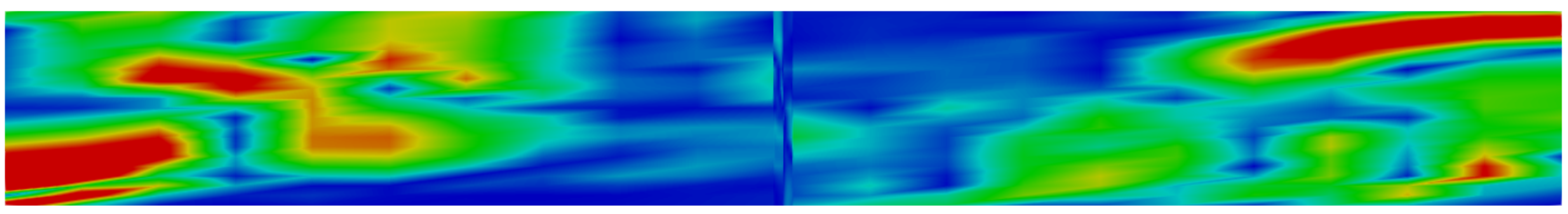}
\end{minipage}%
}%
\subfigure[error from ROM with 8 POD basis functions, t = 85s]{
\begin{minipage}[t]{0.55\linewidth}
\centering
\includegraphics[width = \linewidth]{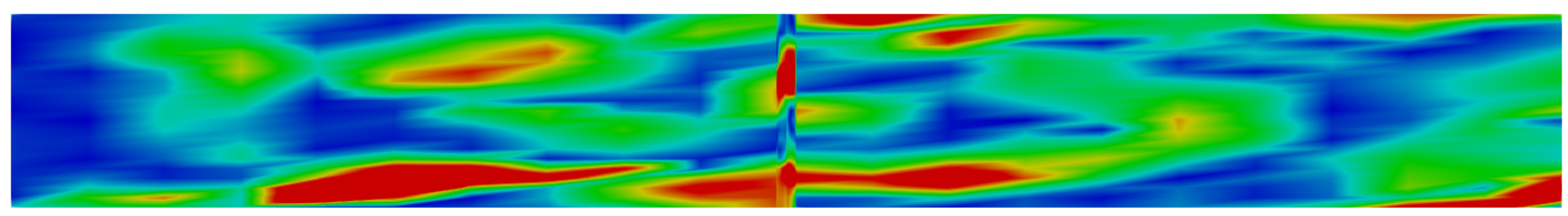}
\end{minipage}%
}%
\\
 
\subfigure[error from DDROM with 5 codes, t = 40s]{
\begin{minipage}[t]{0.55\linewidth}
\centering
\includegraphics[width = \linewidth]{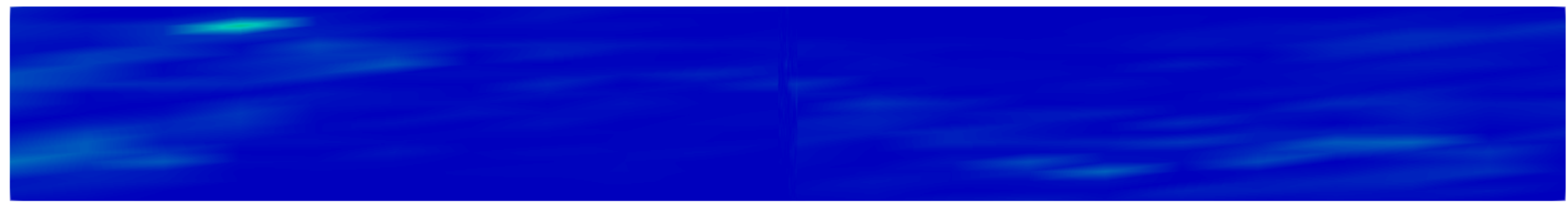}
\end{minipage}%
}%
\subfigure[error from DDROM with 5 codes, t = 85s]{
\begin{minipage}[t]{0.55\linewidth}
\centering
\includegraphics[width = \linewidth]{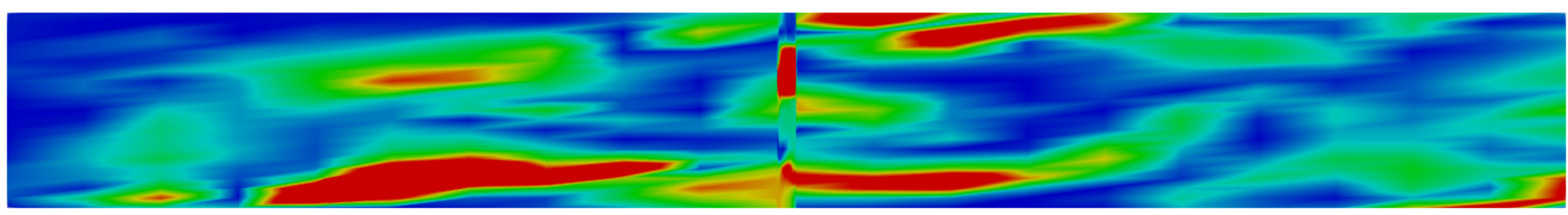}
\end{minipage}%
}%
\\
\subfigure[error from DDROM with 8 codes, t = 40s]{
\begin{minipage}[t]{0.55\linewidth}
\centering
\includegraphics[width = \linewidth]{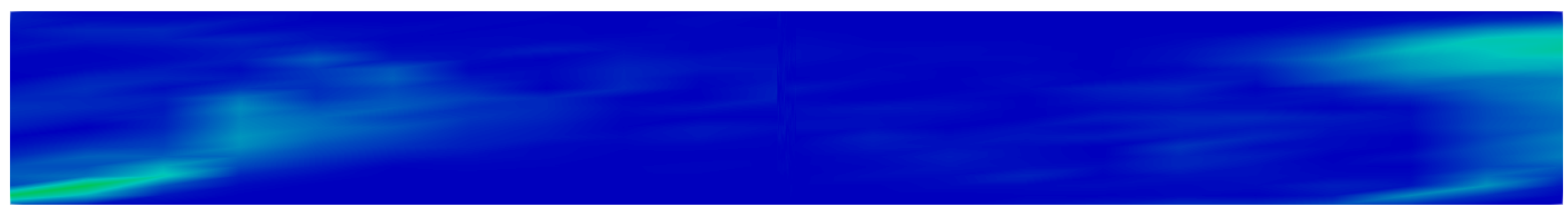}
\end{minipage}%
}%
\subfigure[error from DDROM with 8 codes, t = 85s]{
\begin{minipage}[t]{0.55\linewidth}
\centering
\includegraphics[width = \linewidth]{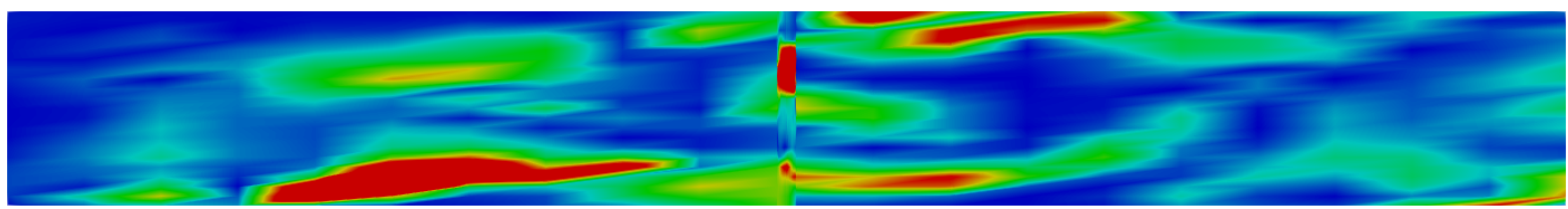}
\end{minipage}%
}%
\\

\begin{minipage}[t]{0.51\linewidth}
\centering
\includegraphics[width = \linewidth]{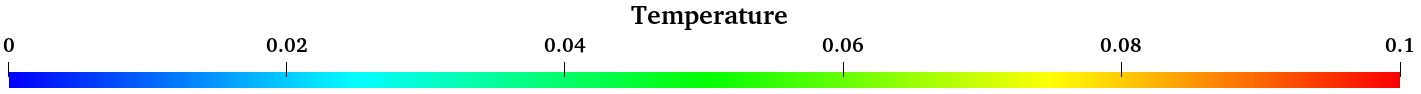}
\end{minipage}%

\begin{minipage}[t]{0.51\linewidth}
\centering
\includegraphics[width = \linewidth]{figs/le_error_legend.png}
\end{minipage}%

\\ 

\end{tabular}
\end{adjustbox}
\caption{Case 1: Lock Exchange. Temperature errors between full model and different ROMs at time levels 40s and 85s. The solutions compare the errors in ROM with 8 POD basis functions(first row) and Auto-Encoder based DDROM with 5 and 8 codes (second and third rows).}
\label{residul-error-lock}

\end{figure}

In order to assess the predictive capability of the model, the predicted solutions at an unseen time level $t=85$s are provided in Figure \ref{lock-predict85}. The temperature solution profiles of these different models are close. In order to see clearly the differences, the temperature solution comparison at a particular point in the computational domain is given in Figure \ref{le_point}.  
The figures show that the DDROM performs very well using as few as 5 codes. In addition, the temperature profile of the DDROM appears to be in closer agreement to the full model solutions than that of POD based ROM. Also, the DDROM with larger number of codes exhibits more accurate solutions. This issue is highlighted in the graphs presented in Figure \ref{lock-cc-rmse}, which show correlation coefficient and the root-mean-square errors (RMSE) of temperature solutions calculated from DDROM with 5 and 8 codes and ROM with 8 POD basis functions. 
Figure \ref{residul-error-lock} shows the residual errors between the high-fidelity full model and the different ROMs at time level t = 40s and t = 85s. As shown in the figure, the overall errors from ROM with 8 POD basis functions are larger than those of Auto-Encoder based DDROM with 5 and 8 codes. The graphs also show that using larger number of codes results in improved accuracy in Auto-Encoder based DDROM. 

\clearpage

\subsection{Case 2: Flow past a cylinder}
\begin{figure}
\centering
\includegraphics[width=0.9\linewidth,angle=0,clip=true]{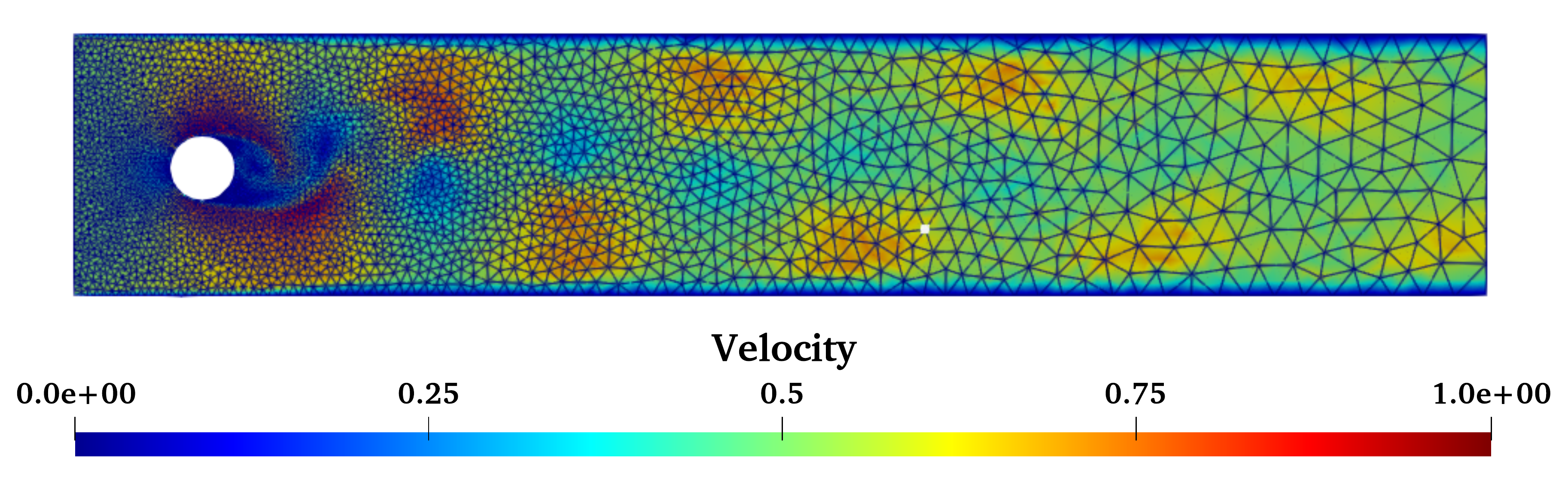}
\caption{Case 2: computational domain of flow past a cylinder test case}
\label{fpc_initial_sta}
\end{figure}

In the second numerical example, a two-dimensional flow past a cylinder is simulated. The computational domain is 2 units in length and 0.4 units in width, and it includes a cylinder of radius 0.12 units positioned over the point (0.2,0.2), as shown in Figure \ref{fpc_initial_sta}. The dynamics of the fluid flow is driven by an in-flowing liquid, that enters the domain via the left boundary with an inlet velocity $V=0.5$. The fluid flows past the cylinder and  the computational domain through the right boundary. No slip and zero outward flow conditions are applied to the lower and upper edges of the domain whilst Dirichlet boundary conditions are applied to the cylinder's wall.  The Reynolds number for this problem is calculated to be $Re = 3200$. 

This problem was simulated for a time period of 200 seconds, with a time step size of $\Delta t = 0.01 $. From the full model simulation, with a mesh of 12568 nodes, 2000 snapshots were obtained at equal time intervals $\Delta t = 0.1 $ for each of the velocity components $(u,v)$ and pressure solution variables. $70\%$ of the simulation data is used for training model. $10\%$ of the simulation data is used to validate the model and $20\%$ of the simulation data is used to test the model. The input number of time levels is 20 in this case.  

\begin{figure}
\centering
\begin{tabular}{cc}

\begin{minipage}{0.5\linewidth} 
\includegraphics[width = \linewidth,angle=0,clip=true]{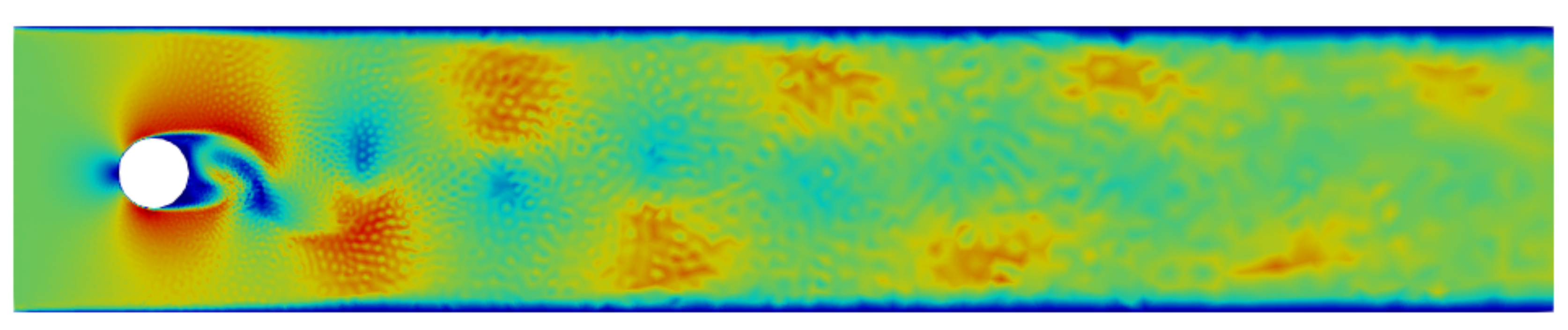}
\end{minipage}
&
\begin{minipage}{0.5\linewidth} 
\includegraphics[width = \linewidth,angle=0,clip=true]{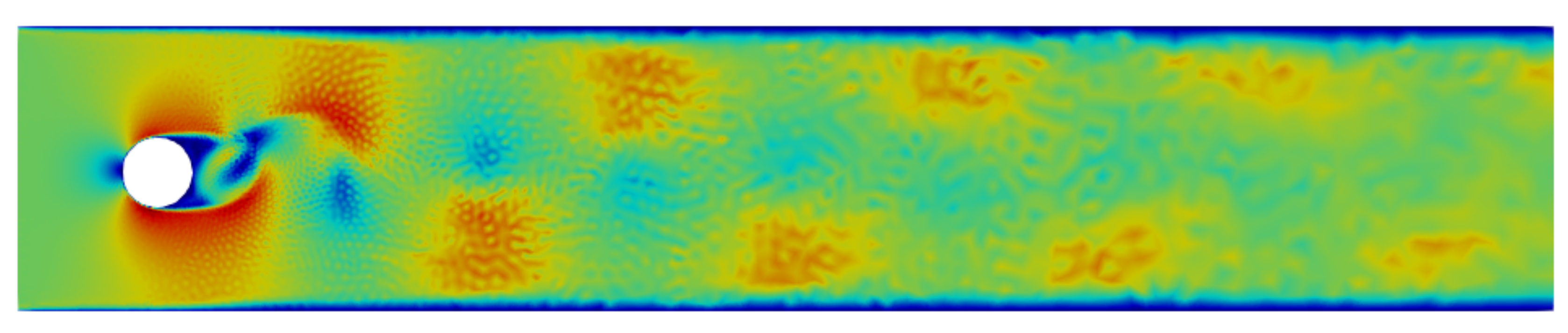}
\end{minipage}\\
(a) {\small full model, $t$ = 100s}&
(b) {\small full model, $t$ = 180s}\\

\begin{minipage}{0.5\linewidth} 
\includegraphics[width = \linewidth,angle=0,clip=true]{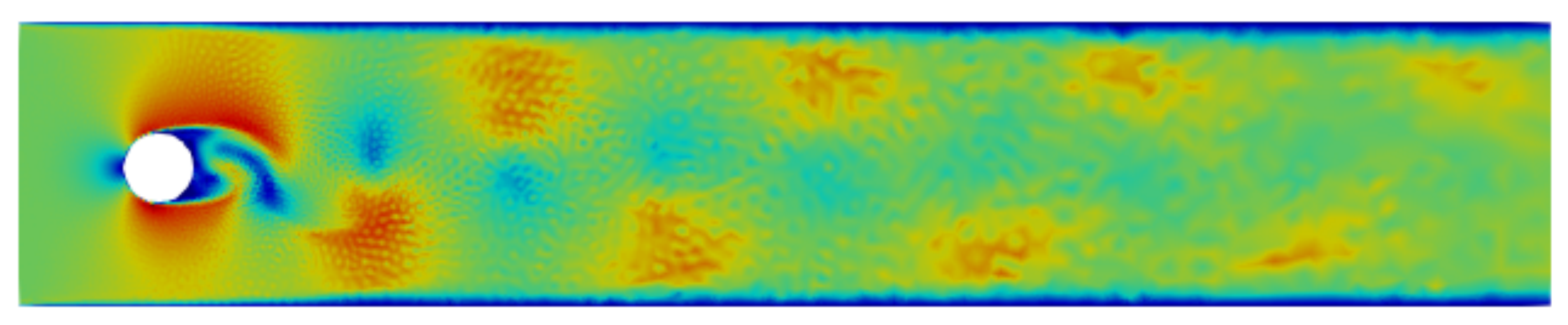}
\end{minipage}
&
\begin{minipage}{0.5\linewidth} 
\includegraphics[width = \linewidth,angle=0,clip=true]{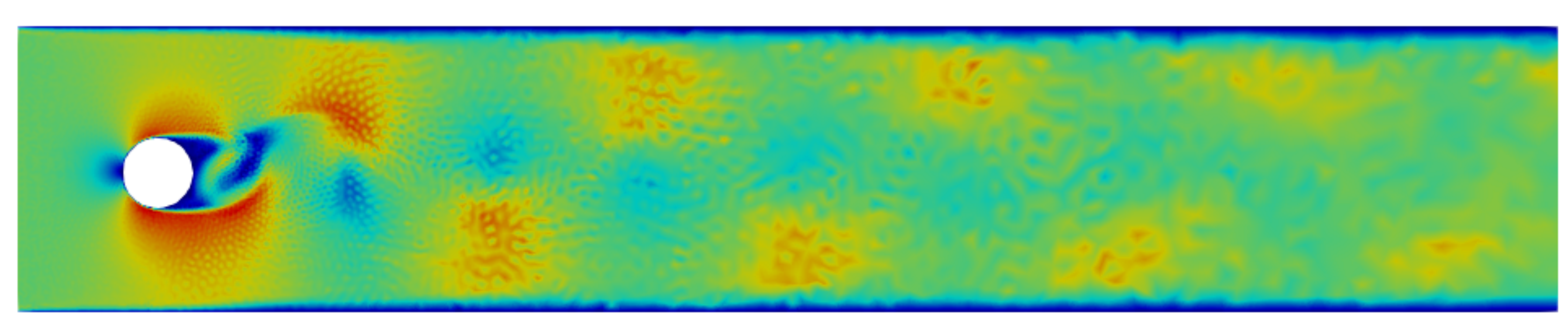}
\end{minipage}\\
(c) {\small DDROM with 3 codes}&
(d) {\small DDROM with 3 codes}\\ 

\begin{minipage}{0.5\linewidth} 
\includegraphics[width = \linewidth,angle=0,clip=true]{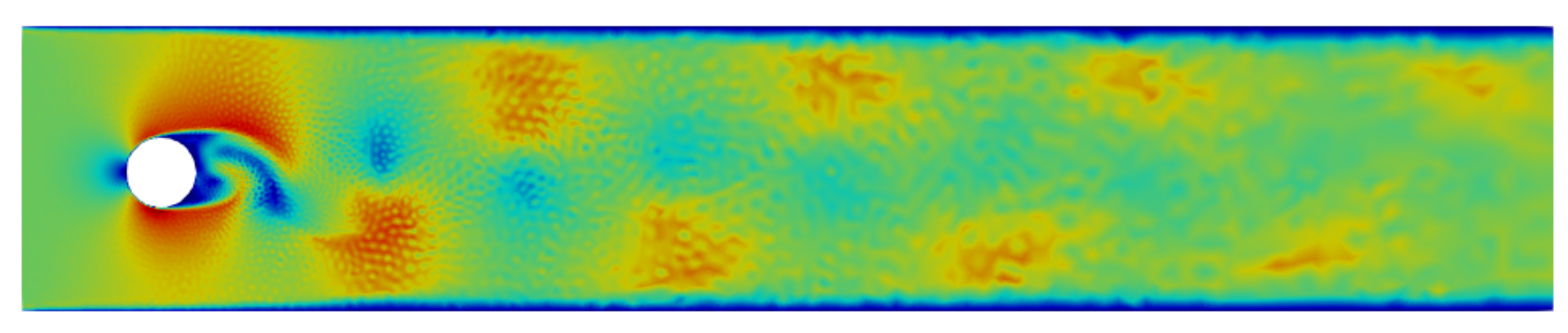}
\end{minipage}
&
\begin{minipage}{0.5\linewidth} 
\includegraphics[width = \linewidth,angle=0,clip=true]{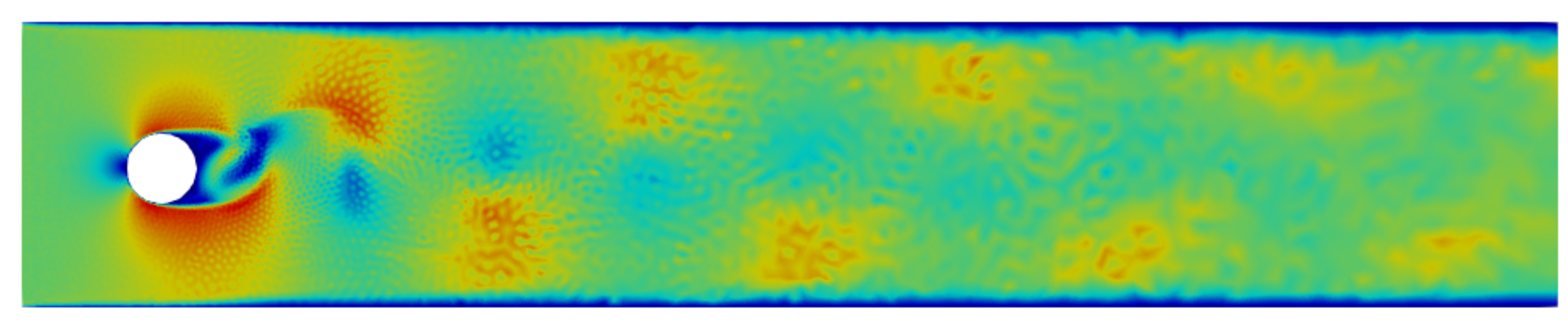}
\end{minipage}\\
(e) {\small DDROM with 4 codes}&
(f) {\small DDROM with 4 codes}\\ 

\begin{minipage}{0.5\linewidth} 
\includegraphics[width = \linewidth,angle=0,clip=true]{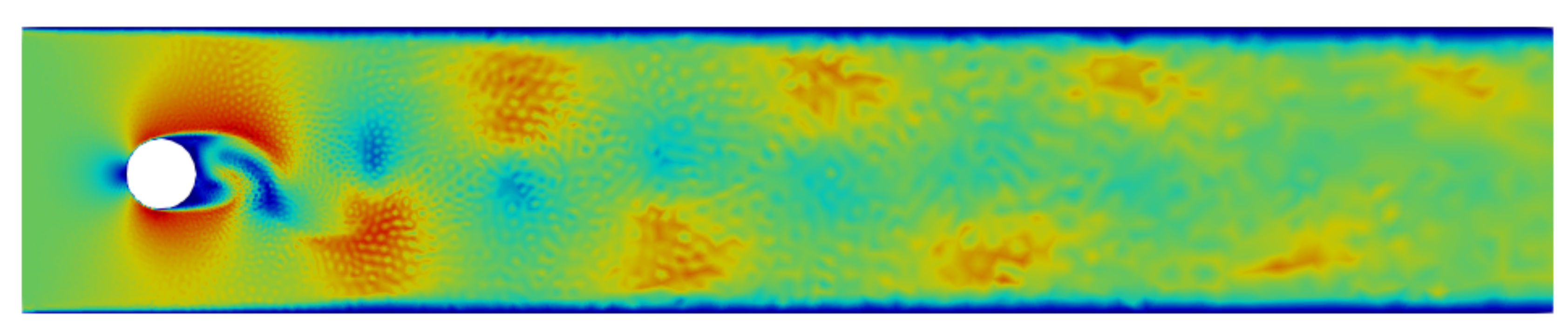}
\end{minipage}
&
\begin{minipage}{0.5\linewidth} 
\includegraphics[width = \linewidth,angle=0,clip=true]{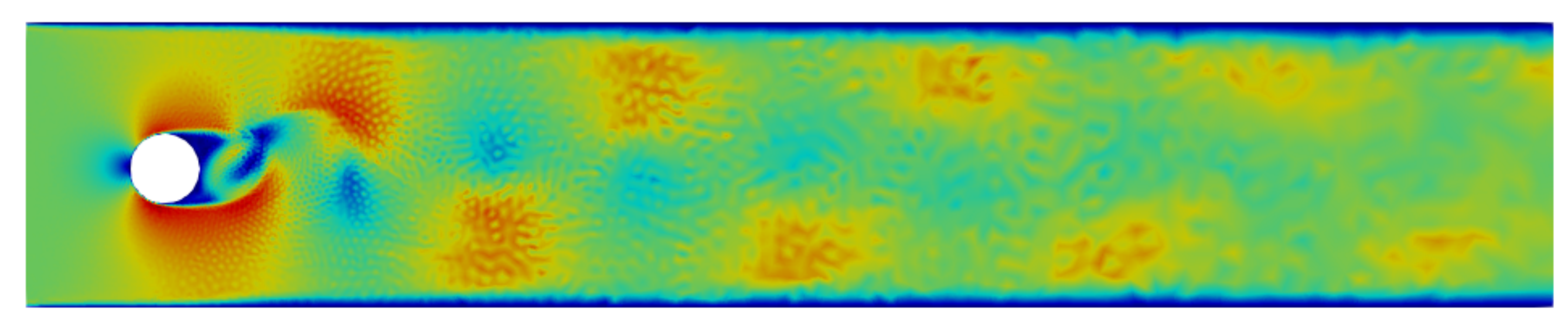}
\end{minipage}\\
(e) {\small DDROM with 6 codes}&
(f) {\small DDROM with 6 codes}\\

\begin{minipage}{0.5\linewidth} 
\includegraphics[width = \linewidth,angle=0,clip=true]{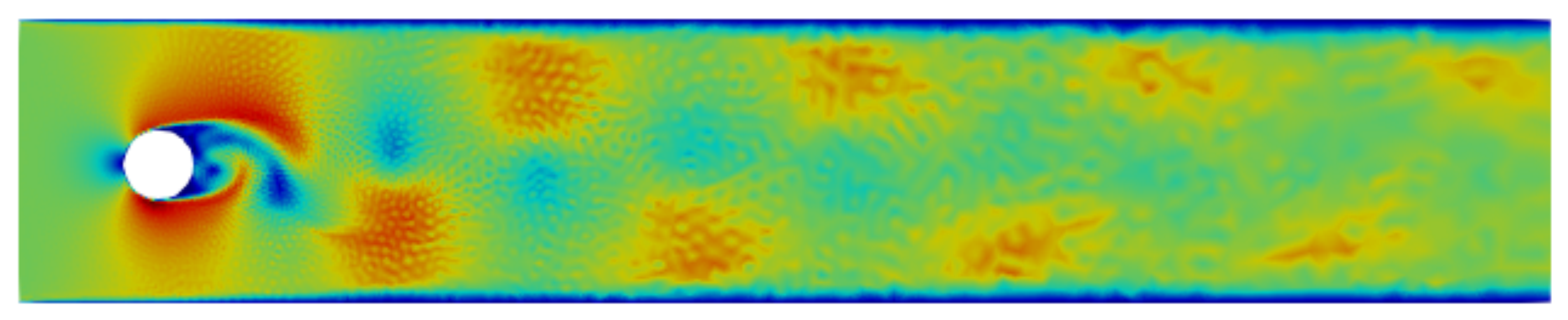}
\end{minipage}
&
\begin{minipage}{0.5\linewidth} 
\includegraphics[width = \linewidth,angle=0,clip=true]{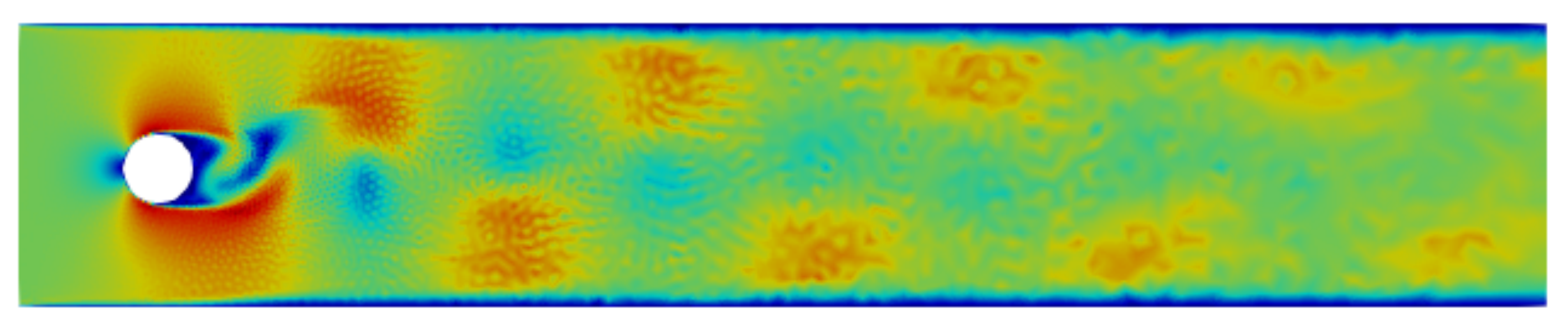}
\end{minipage}\\
(g) {\small ROM with 6 POD basis functions}&
(h) {\small ROM with 6 POD basis functions}\\ 

\begin{minipage}{0.5\linewidth} 
\includegraphics[width = \linewidth,angle=0,clip=true]{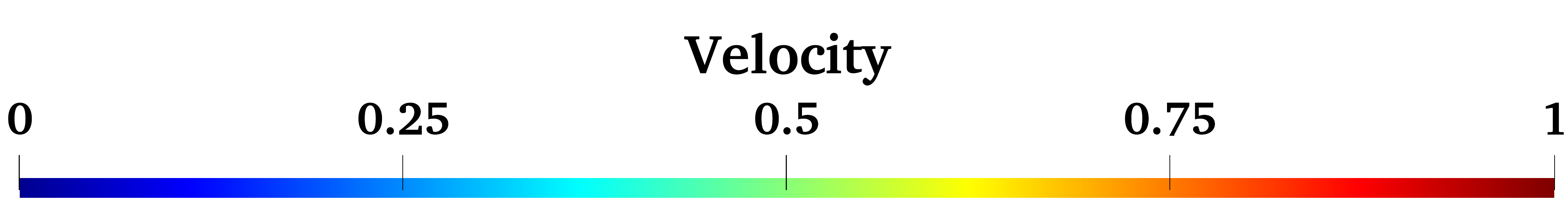}
\end{minipage}
&
\begin{minipage}{0.5\linewidth} 
\includegraphics[width = \linewidth,angle=0,clip=true]{fpc_legend}
\end{minipage}\\
\\ 

\end{tabular}
\caption{Case 2: flow past a cylinder. The graphs (a)-(h) show the field of velocity solutions obtained from the full model, DDROM with 3, 4 and 6 codes and POD based ROM at trained time level $t=100s$ and predicted time level $t=180s$.} 
\label{fpc_over_time}
\end{figure}

\begin{figure}
\centering
\includegraphics[width = \linewidth,angle=0,clip=true]{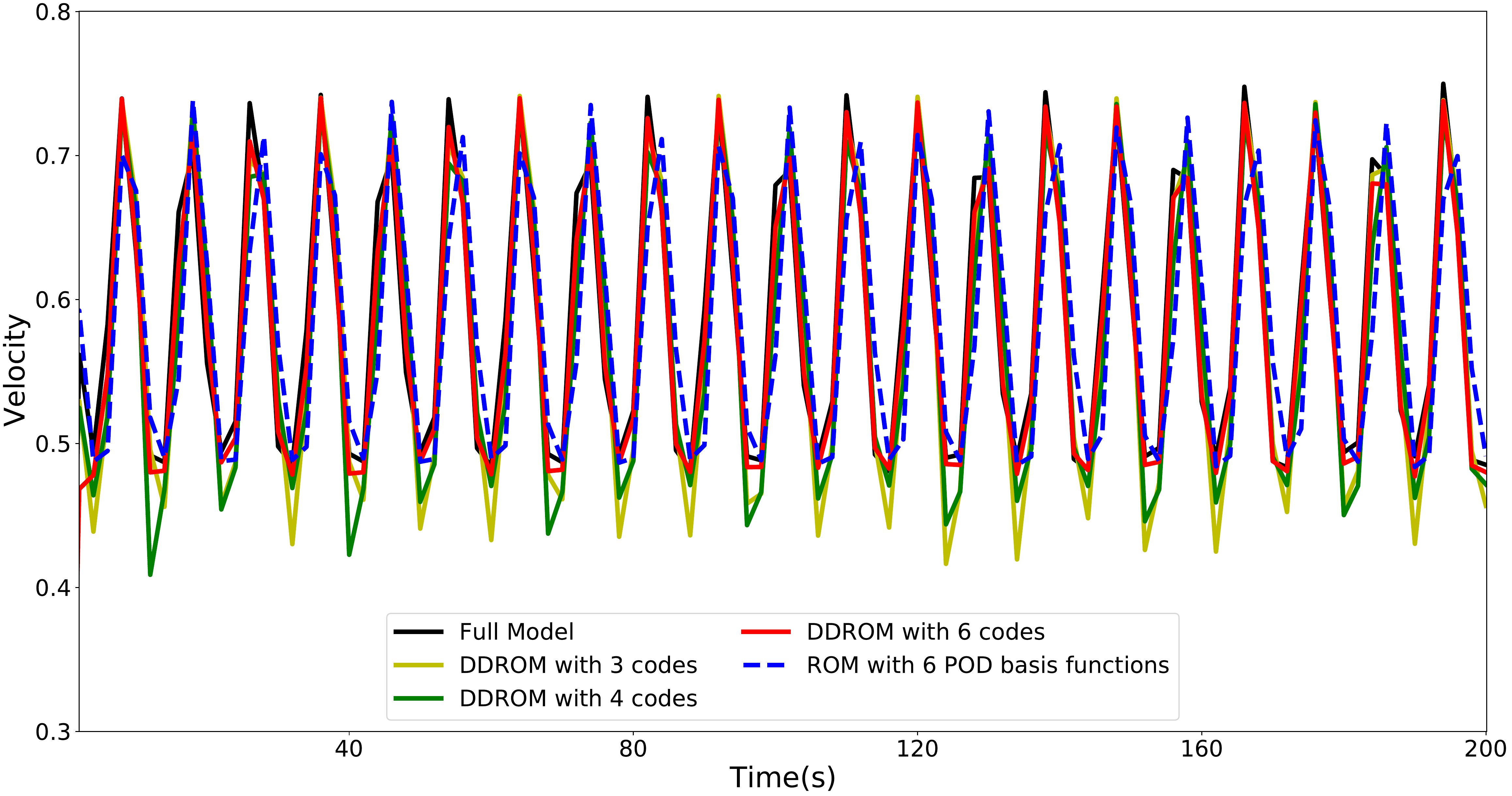}  
\caption{Case 2: flow past a cylinder. The magnitude of velocity of full model, DDROM with 3,4 and 6 codes and POD based ROM with 6 basis functions at a particular point $(x,y)=(1.32233, 0.0606466)$ in the computational domain. The location of this point is shown in Figure \ref{fpc_initial_sta} (see the white point).} 
\label{fpc_point}
\end{figure}

Figure \ref{fpc_over_time} shows the field of velocity solutions obtained from the full model, DDROM with 3, 4 and 6 codes and POD based ROM at trained time level $t=100s$ and predicted time level $t=180s$. As shown in the figure, the solutions of DDROM are closer to the high-fidelity full model when a larger number of codes are used.  
From these flow patterns it is shown that both the DDROM and POD based ROM methods can capture main structural details of the solutions. 
It is also shown that the DDROM performs very well using as few as 3 codes. Additionally, the velocity profile of the DDROM appears to be in closer agreement to the full model solutions than that of ROM based on POD. This issue is highlighted in the graphs presented in Figure \ref{fpc_point} which show the velocity solutions at a particular point $(x=1.32233, y=0.0606466)$, see the white point in Figure \ref{fpc_initial_sta}) in the domain.

The error analysis is carried out by considering correlation coefficients and RMSE taking into account all nodes in the domain.

\begin{figure}
\centering
\begin{tabular}{cc}
\begin{minipage}[b]{0.5\textwidth}
\includegraphics[width=1\textwidth]{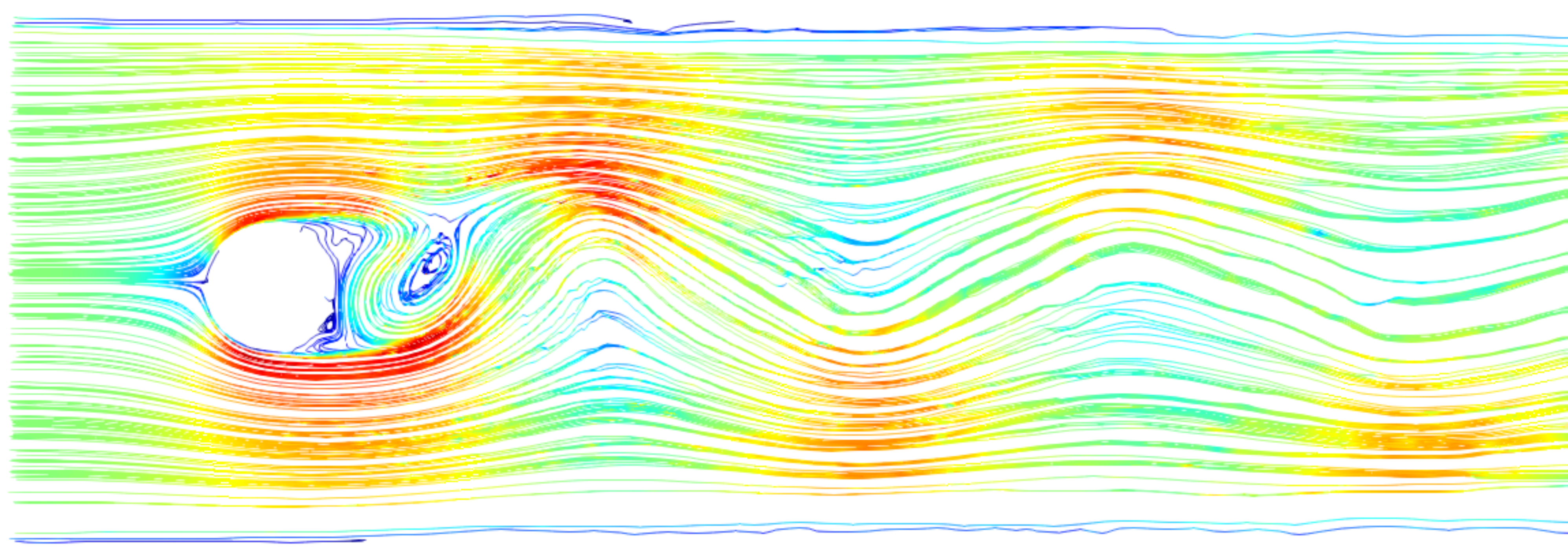}
\end{minipage}
\\
(a) {\small Streamline of Full Model}\\
\begin{minipage}{0.48\linewidth} 
\includegraphics[width = \linewidth,angle=0,clip=true]{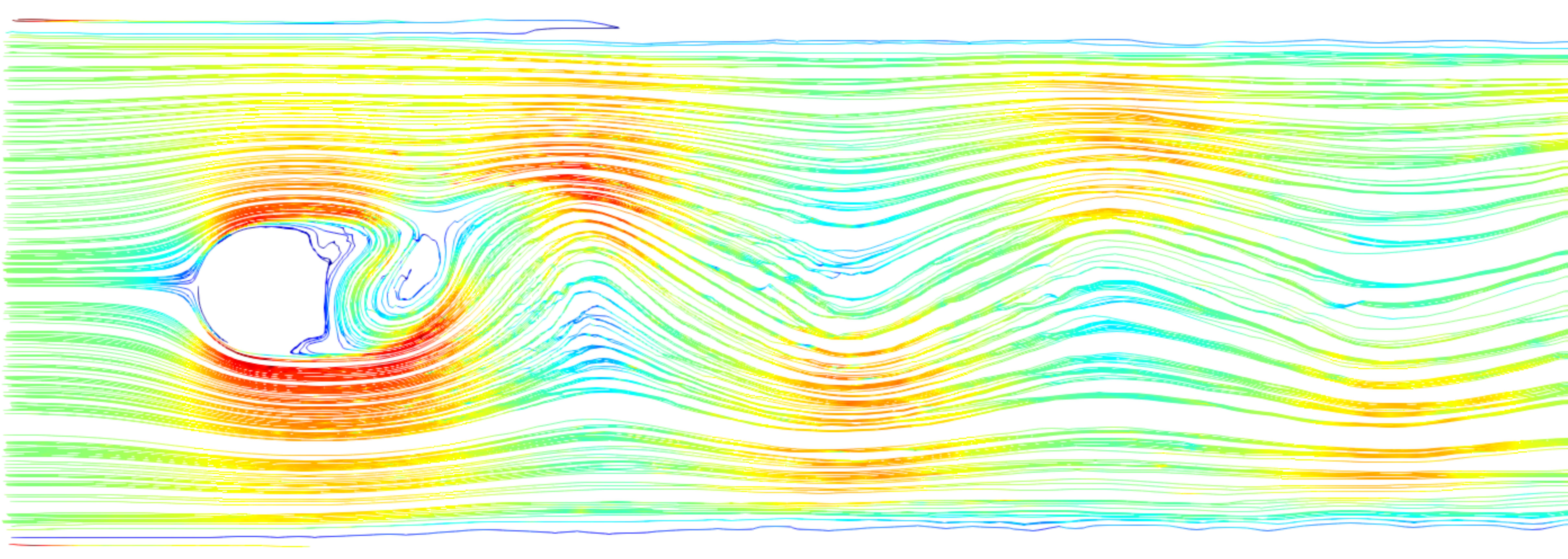}
\end{minipage}
&
\begin{minipage}{0.48\linewidth} 
\includegraphics[width = \linewidth,angle=0,clip=true]{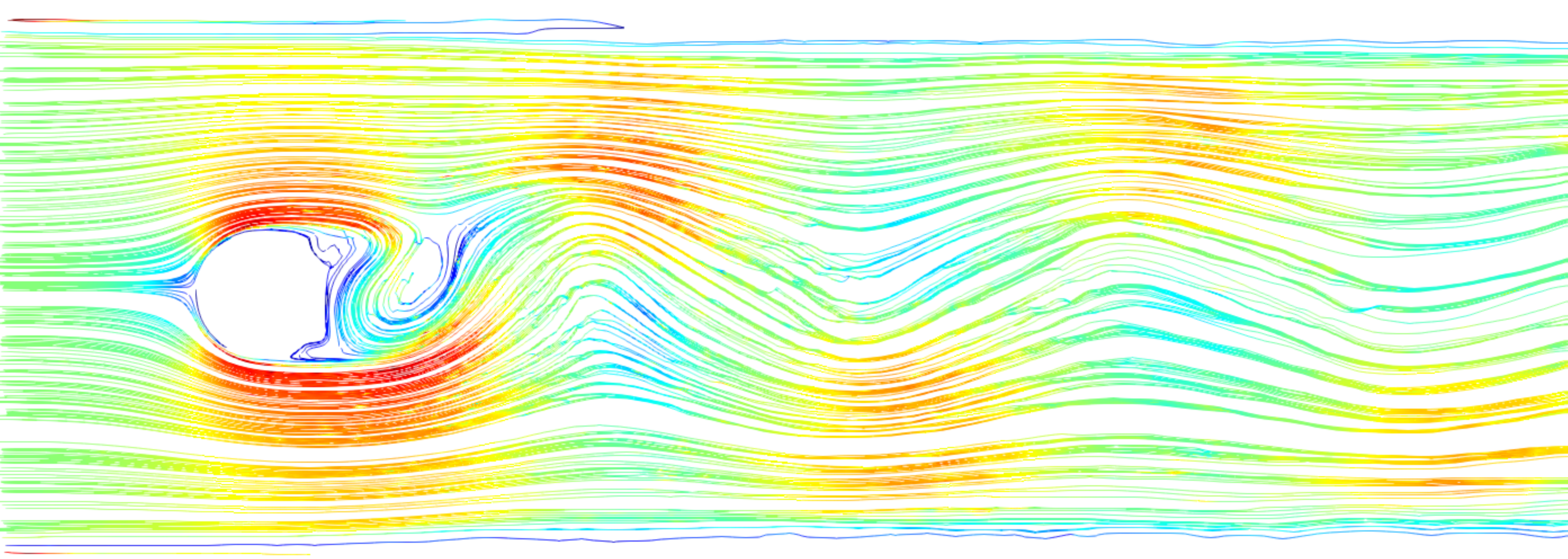}
\end{minipage}\\
(b) {\small Streamline of DDROM with 6 codes}&
(c) {\small Streamline of POD based ROM with 6 basis functions}\\ 
\begin{minipage}{0.5\linewidth} 
\includegraphics[width = \linewidth,angle=0,clip=true]{fpc_legend}
\end{minipage}
&
\begin{minipage}{0.5\linewidth} 
\includegraphics[width = \linewidth,angle=0,clip=true]{fpc_legend}
\end{minipage}\\
\end{tabular}
\caption{Case 2: flow past a cylinder. The streamline of velocity magnitude solutions of full model, DDROM with 6 codes and POD based ROM with 6 basis functions at predicted time level t = 180s. } 
\label{fpc_streamline}
\end{figure}

\begin{figure}
\centering
\begin{tabular}{cc}
\begin{minipage}{0.5\linewidth}
\includegraphics[width = \linewidth,angle=0,clip=true]{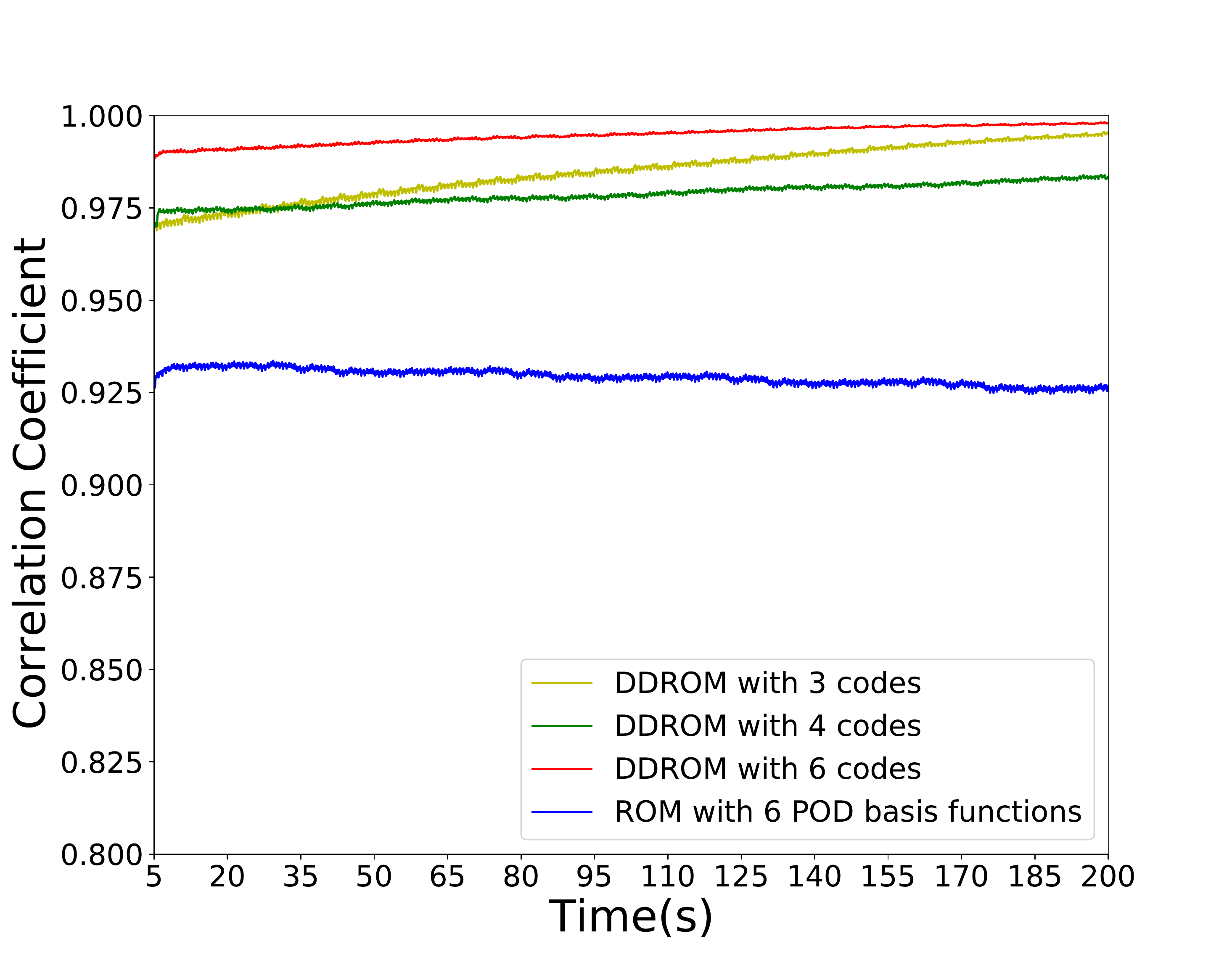}  
\end{minipage}
&
\begin{minipage}{0.5\linewidth}
\includegraphics[width = \linewidth,angle=0,clip=true]{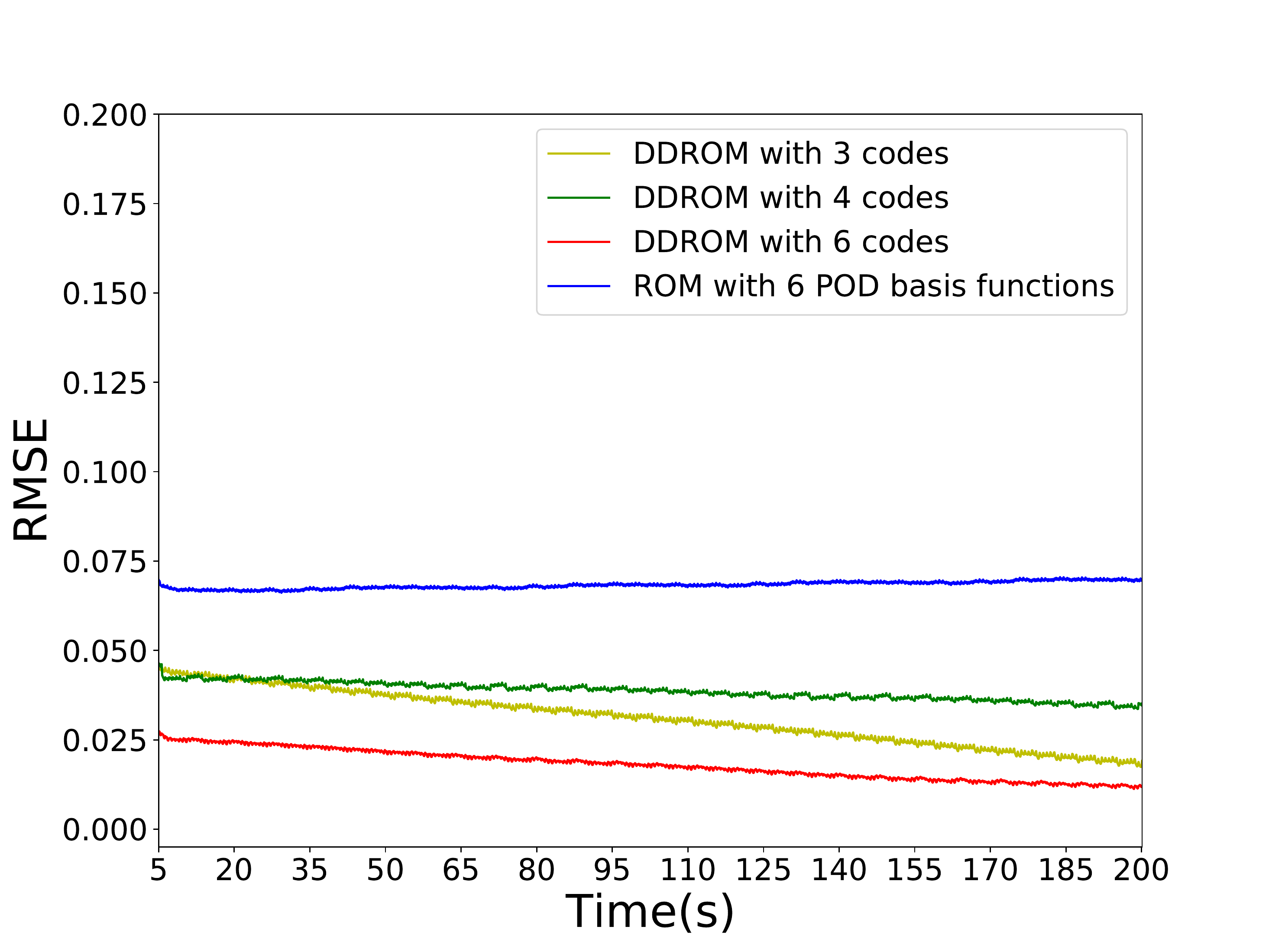}  
\end{minipage} \\
(a) {\small} Correlation Coefficient&
(b) {\small} RMSE\\
\end{tabular}
\caption{Case 2: Flow past a cylinder. Correlation coefficient and the root-mean-square errors(RMSE) of velocity solutions calculated for Auto-Encoder based DDROM with 3, 4, 6 codes and POD based ROM with 6 basis functions.}
\label{flow-cc-rmse}
\end{figure}

\begin{figure}
\centering
\begin{tabular}{cc}
\begin{minipage}{0.48\linewidth}
\includegraphics[width = \linewidth,angle=0,clip=true]{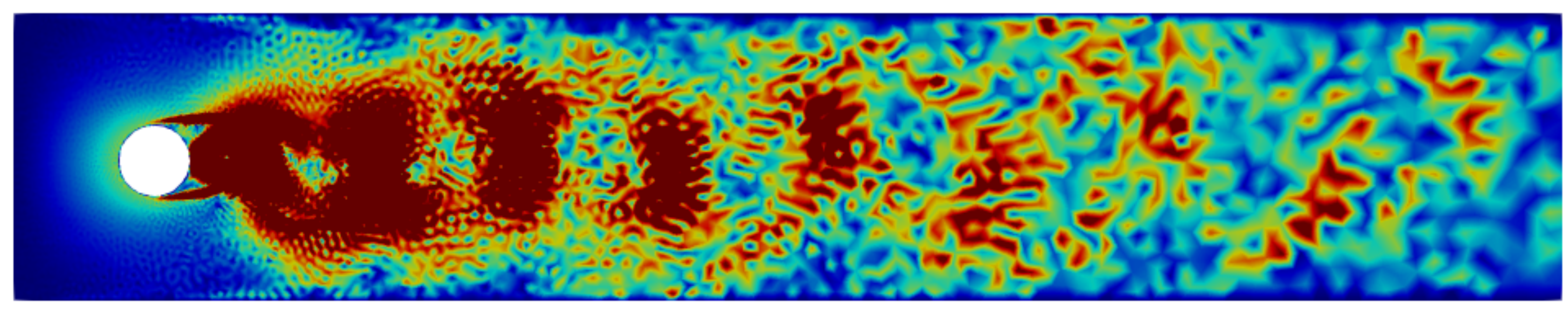}  
\end{minipage}
&
\begin{minipage}{0.48\linewidth}
\includegraphics[width = \linewidth,angle=0,clip=true]{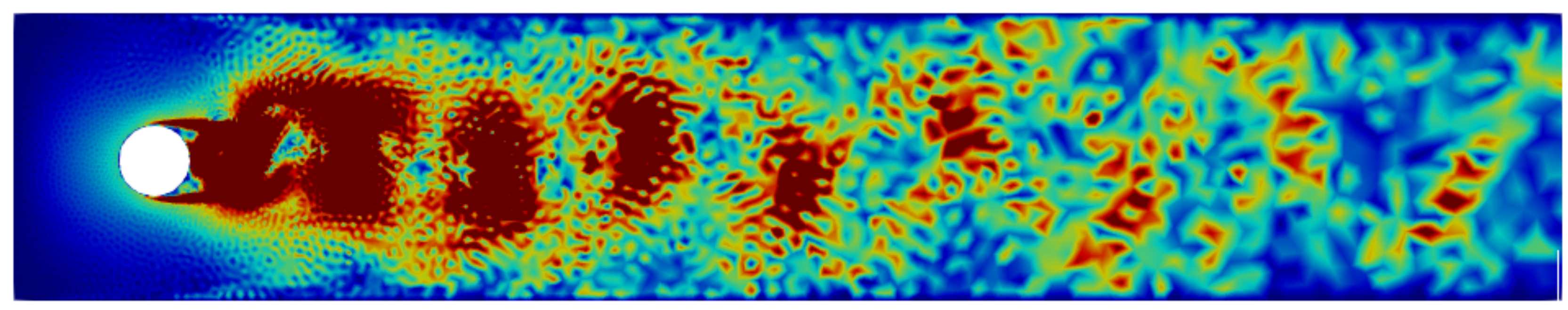}  
\end{minipage} \\
(a) {\small ROM with 6 basis functions error, $t$ = 100s}&
(b) {\small ROM with 6 basis functions error, $t$ = 180s}\\
\begin{minipage}{0.48\linewidth}
\includegraphics[width = \linewidth,angle=0,clip=true]{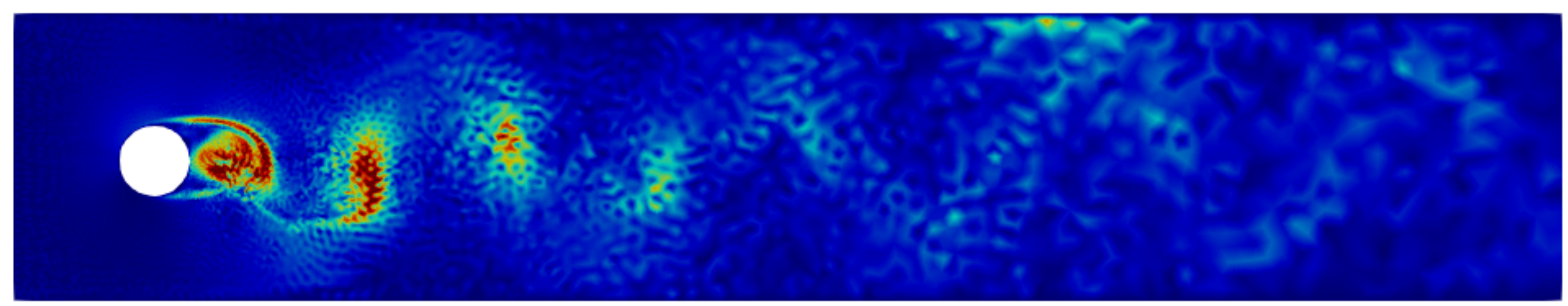}  
\end{minipage}
&
\begin{minipage}{0.48\linewidth}
\includegraphics[width = \linewidth,angle=0,clip=true]{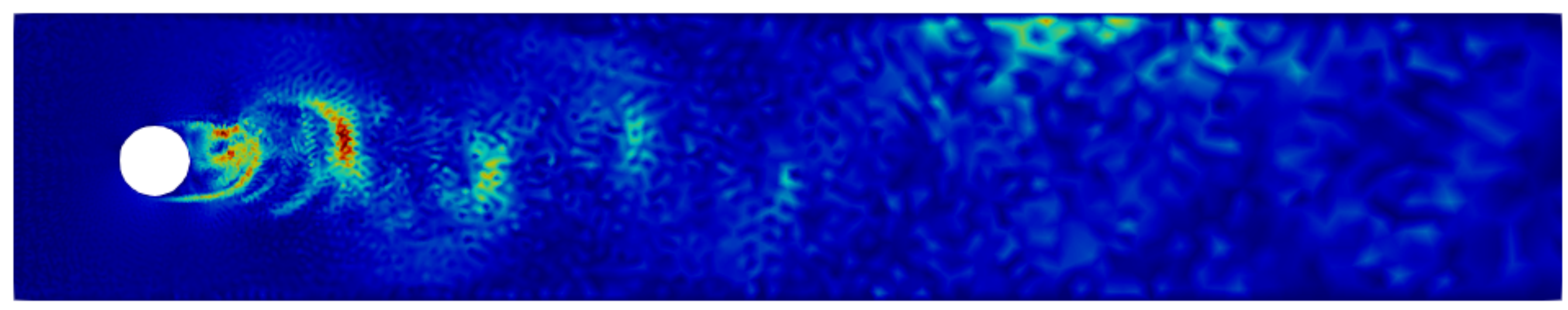}  
\end{minipage} \\
(c) {\small DDROM with 6 codes error, $t$ = 100s}&
(d) {\small DDROM with 6 codes error, $t$ = 180s}\\

\begin{minipage}{0.48\linewidth} 
\includegraphics[width = \linewidth,angle=0,clip=true]{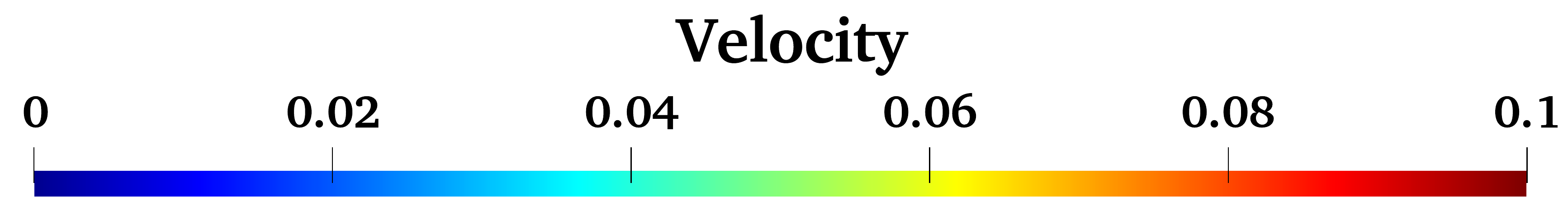}
\end{minipage}
&
\begin{minipage}{0.48\linewidth} 
\includegraphics[width = \linewidth,angle=0,clip=true]{fpc_error_legend}
\end{minipage}\\

\end{tabular}
\caption{Case 2: Flow past a cylinder. Velocity errors of flow past a cylinder problem at time levels 100s(left) and 180s(right). The solutions compare the error in POD based ROM and Auto-Encoder based DDROM. Both models are established using a dimensional size of 6.}
\label{fpc_error}
\end{figure}

The graphs (a) and (b) in Figure \ref{flow-cc-rmse} show errors of the DDROM and POD based ROM and correlation coefficients. Again, these exhibit that a noticeable improvement in accuracy is gained when using the Auto-Encoder network, whereby the errors are reduced in comparison to POD  ROM method. The graphs also show that using larger number of codes results in improved accuracy. The errors between the two ROMs and the high-fidelity full solutions at two time levels t = 100s and t = 180s are presented in Figure \ref{fpc_error}, which shows that the errors decrease using the Auto-Encoder based DDROM. The Auto-Encoder based DDROM is more accurate than that of POD based ROM using identical dimensional size.

\newpage

\subsection{Computational efficiency}
  Table \ref{cputime} shows the CPU cost (online cost and offline cost) required for running the high-fidelity full model and DDROM for each time step. It is worth mentioning that the online CPU cost (dimensionless) required for predicting the DDROM for one time step is only $3.0E-02$ s, while the high-fidelity full model for the flow past a cylinder case is 2.5 s. The computational cost of the high-fidelity full model will increase as the mesh size increases. For a complicated test case with larger number of nodes, the DDROM can gain several orders of magnitude speed up. 
  The training time including SAE and self-attention neural network is big, however, it is offline, which means it does not need to run again after is pre-computed.  
  
  The full model simulations and DDROM online predictions were performed on a workstation with an Intel 8 cores i7-9700 Processor (3.00GHz base frequency and 4.70GHz Max Turbo Frequency) and 24G RAM. The training of DDROM were performed on Google's Colaboratory platform, which allows AI researchers to write and execute python scripts in the web browser. It also provides GPUs for deep learning training.


\begin{table*}[htbp!]
\label{cputime}
  \centering
  \caption{CPU time required to run the full-fidelity full model and DDROM(s)}
  \begin{threeparttable}
    \begin{tabular}{c c c c c c c c } 
      \hline
      Cases    & Models & SAE training\tnote{*} &  Self-attention training\tnote{**} & Prediction &  Nodes \\ \hline
      Lock & full model     & \verb|\| & \verb|\|   & 0.12   &  1490 \\ 
       exchange        & POD(8)      & \verb|\| & 374.2609    & $2.3554E-02$  &  1490 \\
                       & DDROM(5)    & 359.6998 & 326.5726   &  $2.3547E-02$  &  1490 \\
                       & DDROM(8)    & 364.1556 & 446.0504    & $2.2545E-02$  &  1490 \\
      \hline
      Flow       & full model        & \verb|\| & \verb|\|     & 2.5   & 12568  \\ 
      past a     & POD(6)        & \verb|\|     &185.9721     & $3.0387E-02$   & 12568 \\ 
      cylinder   & DDROM(3)      &277.5404       & 214.5255   & $3.0064E-02$   &  12568 \\
                 & DDROM(4)      &282.4974       & 181.5445   & $3.0087E-02$  &  12568 \\
                 & DDROM(6)      &267.5012       & 262.0736   & $3.0296E-02$   &  12568 \\

      \hline
    \end{tabular}
 \begin{tablenotes}
        \footnotesize
        \item[*]  The training time of Auto-Encoder model for 300 epochs  
        \item[**] The training time of Self-attention model for 100 epochs  
      \end{tablenotes}
    \end{threeparttable}
    
\end{table*}

\clearpage
\section{Conclusions}\label{Conclusions}
A new data-driven non-intrusive reduced order model (DDROM) is presented in this work. This is achieved by constructing a new model reduction deep learning neural network architecture. The architecture consists of two main parts: Auto-Encoder part and self-attention part. The Auto-Encoder part is used to project the high-dimensional full space into a much lower dimensional space while the self-attention part involves constructing a number of functions representing the fluid dynamics in reduced space. The new DDROM has been implemented under the framework of an advanced three dimensional finite element mesh fluid model (Fluidity).  The performance of this DDROM has been illustrated by two numerical problems: flow past a cylinder and a lock exchange test cases. A detailed comparison between the high-fidelity full model and the DDROM has been made. The solutions show that Auto-Encoder based DDROM performs better than POD based ROM. A significant CPU speed-up is obtained in comparison to the high-fidelity full model. The advantage of this DDROM is that it is able to capture more non-linearity information than POD based ROM. 

Future work will apply this DDROM into more complicated problems such as large scale urban flows, flooding problems, multi-phase problems and ocean problems. The predictive capability of DDROM has strong connections with the amount of training data. Larger number of training data will lead to higher accuracy while it is also case dependent. Future work will also include developing a parametric DDROM which is able to deal with varying initial and boundary conditions.

\vspace{-6pt}
 
\section*{Acknowledgments}
\noindent 
The authors would like to acknowledge the support of EPSRC grant: PURIFY ($EP/V000756/1$) and the Royal Society International Exchanges 2019 Cost Share ($IEC \verb|\| NSFC \verb|\|191037$). Rui Fu acknowledges College of Engineering Centenary PhD Scholarship at Swansea University. We acknowledge the support of the Supercomputing Wales project, which is part-funded by the European Regional Development Fund (ERDF) via Welsh Government.

\clearpage

\bibliographystyle{unsrt} 
\bibliography{bibliography}

\begin{thebibliography}{10}

\bibitem{hijazi2020data}
Saddam Hijazi, Giovanni Stabile, Andrea Mola, and Gianluigi Rozza.
\newblock Data-driven {POD-Galerkin} reduced order model for turbulent flows.
\newblock {\em Journal of Computational Physics}, 416:109513, 2020.

\bibitem{rudy2017data}
Samuel~H Rudy, Steven~L Brunton, Joshua~L Proctor, and J~Nathan Kutz.
\newblock Data-driven discovery of partial differential equations.
\newblock {\em Science Advances}, 3(4):e1602614, 2017.

\bibitem{vermeulen2006model}
PTM Vermeulen and AW~Heemink.
\newblock Model-reduced variational data assimilation.
\newblock {\em {}Monthly weather review}.

\bibitem{cao2006reduced}
Yanhua Cao, Jiang Zhu, Zhendong Luo, and Ionel~M Navon.
\newblock Reduced-order modeling of the upper tropical pacific ocean model
  using proper orthogonal decomposition.
\newblock {\em Computers \& mathematics with Applications}, 52(8-9):1373--1386,
  2006.

\bibitem{daescu2008dual}
DN~Daescu and IM~Navon.
\newblock A dual-weighted approach to order reduction in 4dvar data
  assimilation.
\newblock {\em Monthly Weather Review}, 136(3):1026--1041, 2008.

\bibitem{cstefuanescu2013pod}
R{\u{a}}zvan {\c{S}}tef{\u{a}}nescu and Ionel~Michael Navon.
\newblock {POD/DEIM} nonlinear model order reduction of an adi implicit shallow
  water equations model.
\newblock {\em Journal of Computational Physics}, 237:95--114, 2013.

\bibitem{XIAO2019463}
D.~Xiao, F.~Fang, J.~Zheng, C.C. Pain, and I.M. Navon.
\newblock Machine learning-based rapid response tools for regional air
  pollution modelling.
\newblock {\em Atmospheric Environment}, 199:463--473, 2019.

\bibitem{benner2021interpolation}
Peter Benner and Pawan Goyal.
\newblock Interpolation-based model order reduction for polynomial systems.
\newblock {\em SIAM Journal on Scientific Computing}, 43(1):A84--A108, 2021.

\bibitem{martini2018certified}
Immanuel Martini, Bernard Haasdonk, and Gianluigi Rozza.
\newblock Certified reduced basis approximation for the coupling of viscous and
  inviscid parametrized flow models.
\newblock {\em Journal of Scientific Computing}, 74(1):197--219, 2018.

\bibitem{peherstorfer2017data}
Benjamin Peherstorfer, Serkan Gugercin, and Karen Willcox.
\newblock Data-driven reduced model construction with time-domain loewner
  models.
\newblock {\em SIAM Journal on Scientific Computing}, 39(5):A2152--A2178, 2017.

\bibitem{alla2018posteriori}
Alessandro Alla, Carmen Gr{\"a}ssle, and Michael Hinze.
\newblock A posteriori snapshot location for {POD} in optimal control of linear
  parabolic equations.
\newblock {\em ESAIM: Mathematical Modelling and Numerical Analysis},
  52(5):1847--1873, 2018.

\bibitem{son2015model}
Nguyen~Thanh Son and Tatjana Stykel.
\newblock Model order reduction of parameterized circuit equations based on
  interpolation.
\newblock {\em Advances in Computational Mathematics}, 41(5):1321--1342, 2015.

\bibitem{hachtel2018multirate}
Christoph Hachtel, Johanna Kerler-Back, Andreas Bartel, Michael G{\"u}nther,
  and Tatjana Stykel.
\newblock Multirate dae/ode-simulation and model order reduction for coupled
  field-circuit systems.
\newblock In {\em Scientific Computing in Electrical Engineering}, pages
  91--100. Springer, 2018.

\bibitem{cstefuanescu2016model}
R{\u{a}}zvan {\c{S}}tef{\u{a}}nescu, Bernd~R Noack, and Adrian Sandu.
\newblock Model reduction and inverse problems and data assimilation with
  geophysical applications. a special issue in honor of i. michael navon's 75th
  birthday, 2016.

\bibitem{xie2018data}
Xuping Xie, Muhammad Mohebujjaman, Leo~G Rebholz, and Traian Iliescu.
\newblock Data-driven filtered reduced order modeling of fluid flows.
\newblock {\em {SIAM Journal on Scientific Computing}}, 40(3):B834--B857, 2018.

\bibitem{xiao2021efficient}
Cong Xiao, Olwijn Leeuwenburgh, Hai~Xiang Lin, and Arnold Heemink.
\newblock Efficient estimation of space varying parameters in numerical models
  using non-intrusive subdomain reduced order modeling.
\newblock {\em Journal of Computational Physics}, 424:109867, 2021.

\bibitem{carlberg2013gnat}
Kevin Carlberg, Charbel Farhat, Julien Cortial, and David Amsallem.
\newblock The gnat method for nonlinear model reduction: effective
  implementation and application to computational fluid dynamics and turbulent
  flows.
\newblock {\em Journal of Computational Physics}, 242:623--647, 2013.

\bibitem{FLM9529305}
Michael Schlegel and Bernd~R. Noack.
\newblock {On long-term boundedness of Galerkin models}.
\newblock {\em Journal of Fluid Mechanics}, 765:325--352, 2 2015.

\bibitem{FLM9241738}
Jan Osth, Bernd~R. Noack, Sinisa Krajnovic, Diogo Barros, and Jacques Boree.
\newblock {On the need for a nonlinear subscale turbulence term in {POD} models
  as exemplified for a high-Reynolds-number flow over an Ahmed body}.
\newblock {\em Journal of Fluid Mechanics}, 747:518--544, 5 2014.

\bibitem{Chaturantabut2010}
S.~Chaturantabut and D.C. Sorensen.
\newblock Nonlinear model reduction via discrete empirical interpolation.
\newblock {\em SIAM J. Sci. Comput}, 32:2737--2764, 2010.

\bibitem{carlberg2011efficient}
Kevin Carlberg, Charbel Bou-Mosleh, and Charbel Farhat.
\newblock Efficient non-linear model reduction via a least-squares
  {Petrov--Galerkin} projection and compressive tensor approximations.
\newblock {\em International Journal for numerical methods in engineering},
  86(2):155--181, 2011.

\bibitem{willcox2003model}
Karen Willcox and Alexandre Megretski.
\newblock Model reduction for large-scale linear applications.
\newblock {\em IFAC Proceedings Volumes}, 36(16):1387--1392, 2003.

\bibitem{xiao2013non}
D~Xiao, F~Fang, J~Du, CC~Pain, IM~Navon, AG~Buchan, Ahmed~H Elsheikh, and G~Hu.
\newblock Non-linear {Petrov--Galerkin} methods for reduced order modelling of
  the {Navier--Stokes} equations using a mixed finite element pair.
\newblock {\em Computer Methods In Applied Mechanics and Engineering},
  255:147--157, 2013.

\bibitem{barrault2004empirical}
Maxime Barrault, Yvon Maday, Ngoc~Cuong Nguyen, and Anthony~T Patera.
\newblock An ‘empirical interpolation’method: application to efficient
  reduced-basis discretization of partial differential equations.
\newblock {\em Comptes Rendus Mathematique}, 339(9):667--672, 2004.

\bibitem{chaturantabut2011application}
Saifon Chaturantabut and Danny~C Sorensen.
\newblock Application of {POD} and {DEIM} on dimension reduction of non-linear
  miscible viscous fingering in porous media.
\newblock {\em Mathematical and Computer Modelling of Dynamical Systems},
  17(4):337--353, 2011.

\bibitem{chaturantabut2012state}
Saifon Chaturantabut and Danny~C Sorensen.
\newblock A state space error estimate for {POD-DEIM} nonlinear model
  reduction.
\newblock {\em {SIAM Journal on numerical analysis}}, 50(1):46--63, 2012.

\bibitem{xiao2014non}
Dunhui Xiao, Fangxin Fang, Andrew~G Buchan, Christopher~C Pain, Ionel~Michael
  Navon, Juan Du, and G~Hu.
\newblock Non-linear model reduction for the {Navier--Stokes} equations using
  residual {DEIM} method.
\newblock {\em Journal of Computational Physics}, 263:1--18, 2014.

\bibitem{xiao2015non}
D~Xiao, F~Fang, AG~Buchan, CC~Pain, IM~Navon, and A~Muggeridge.
\newblock Non-intrusive reduced order modelling of the {Navier--Stokes}
  equations.
\newblock {\em Computer Methods in Applied Mechanics and Engineering},
  293:522--541, 2015.

\bibitem{peherstorfer2016data}
Benjamin Peherstorfer and Karen Willcox.
\newblock Data-driven operator inference for nonintrusive projection-based
  model reduction.
\newblock {\em Computer Methods in Applied Mechanics and Engineering},
  306:196--215, 2016.

\bibitem{rahman2019nonintrusive}
Sk~M Rahman, Suraj Pawar, Omer San, Adil Rasheed, and Traian Iliescu.
\newblock Nonintrusive reduced order modeling framework for quasigeostrophic
  turbulence.
\newblock {\em Physical Review E}, 100(5):053306, 2019.

\bibitem{kast2020non}
Mariella Kast, Mengwu Guo, and Jan~S Hesthaven.
\newblock A non-intrusive multifidelity method for the reduced order modeling
  of nonlinear problems.
\newblock {\em Computer Methods in Applied Mechanics and Engineering},
  364:112947, 2020.

\bibitem{mou2021data}
Changhong Mou, Birgul Koc, Omer San, Leo~G Rebholz, and Traian Iliescu.
\newblock Data-driven variational multiscale reduced order models.
\newblock {\em Computer Methods in Applied Mechanics and Engineering},
  373:113470, 2021.

\bibitem{xiao2019domain}
Dunhui Xiao, Fangxin Fang, Claire~E Heaney, IM~Navon, and CC~Pain.
\newblock A domain decomposition method for the non-intrusive reduced order
  modelling of fluid flow.
\newblock {\em Computer Methods in Applied Mechanics and Engineering},
  354:307--330, 2019.

\bibitem{wang2019non}
Qian Wang, Jan~S Hesthaven, and Deep Ray.
\newblock Non-intrusive reduced order modeling of unsteady flows using
  artificial neural networks with application to a combustion problem.
\newblock {\em Journal of computational physics}, 384:289--307, 2019.

\bibitem{jacquier2021non}
Pierre Jacquier, Azzedine Abdedou, Vincent Delmas, and Azzeddine
  Soula{\"\i}mani.
\newblock Non-intrusive reduced-order modeling using uncertainty-aware deep
  neural networks and proper orthogonal decomposition: Application to flood
  modeling.
\newblock {\em Journal of Computational Physics}, 424:109854, 2021.

\bibitem{ahmed2019memory}
Shady~E Ahmed, Sk~Mashfiqur Rahman, Omer San, Adil Rasheed, and Ionel~M Navon.
\newblock Memory embedded non-intrusive reduced order modeling of non-ergodic
  flows.
\newblock {\em Physics of Fluids}, 31(12):126602, 2019.

\bibitem{lee2020model}
Kookjin Lee and Kevin~T Carlberg.
\newblock Model reduction of dynamical systems on nonlinear manifolds using
  deep convolutional autoencoders.
\newblock {\em Journal of Computational Physics}, 404:108973, 2020.

\bibitem{kurschner2018balanced}
Patrick K{\"u}rschner.
\newblock Balanced truncation model order reduction in limited time intervals
  for large systems.
\newblock {\em Advances in Computational Mathematics}, 44(6):1821--1844, 2018.

\bibitem{gugercin2008h_2}
Serkan Gugercin, Athanasios~C Antoulas, and Christopher Beattie.
\newblock H\_2 model reduction for large-scale linear dynamical systems.
\newblock {\em SIAM journal on matrix analysis and applications},
  30(2):609--638, 2008.

\bibitem{pagani2018numerical}
Stefano Pagani, Andrea Manzoni, and Alfio Quarteroni.
\newblock Numerical approximation of parametrized problems in cardiac
  electrophysiology by a local reduced basis method.
\newblock {\em Computer Methods in Applied Mechanics and Engineering},
  340:530--558, 2018.

\bibitem{lecun2015deep}
Yann LeCun, Yoshua Bengio, and Geoffrey Hinton.
\newblock Deep learning.
\newblock {\em nature}, 521(7553):436--444, 2015.

\bibitem{vamathevan2019applications}
Jessica Vamathevan, Dominic Clark, Paul Czodrowski, Ian Dunham, Edgardo Ferran,
  George Lee, Bin Li, Anant Madabhushi, Parantu Shah, Michaela Spitzer, et~al.
\newblock Applications of machine learning in drug discovery and development.
\newblock {\em Nature Reviews Drug Discovery}, 18(6):463--477, 2019.

\bibitem{tagade2019attribute}
Piyush~M Tagade, Shashishekar~P Adiga, Shanthi Pandian, Min~Sik Park,
  Krishnan~S Hariharan, and Subramanya~Mayya Kolake.
\newblock Attribute driven inverse materials design using deep learning
  bayesian framework.
\newblock {\em npj Computational Materials}, 5(1):1--14, 2019.

\bibitem{tran2021fast}
Duc Tran, Hung Nguyen, Bang Tran, Carlo La~Vecchia, Hung~N Luu, and Tin Nguyen.
\newblock Fast and precise single-cell data analysis using a hierarchical
  autoencoder.
\newblock {\em Nature communications}, 12(1):1--10, 2021.

\bibitem{phillips2021autoencoder}
Toby~RF Phillips, Claire~E Heaney, Paul~N Smith, and Christopher~C Pain.
\newblock An autoencoder-based reduced-order model for eigenvalue problems with
  application to neutron diffusion.
\newblock {\em International Journal for Numerical Methods in Engineering},
  122(15):3780--3811, 2021.

\bibitem{wu2021reduced}
Pin Wu, Siquan Gong, Kaikai Pan, Feng Qiu, Weibing Feng, and Christopher Pain.
\newblock Reduced order model using convolutional auto-encoder with
  self-attention.
\newblock {\em Physics of Fluids}, 33(7):077107, 2021.

\bibitem{waswani2017attention}
A~Waswani, N~Shazeer, N~Parmar, J~Uszkoreit, L~Jones, AN~Gomez, L~Kaiser, and
  I~Polosukhin.
\newblock Attention is all you need.
\newblock In {\em NIPS}, 2017.

\bibitem{phan2021self}
Huy Phan, Huy Le~Nguyen, Oliver~Y Ch{\'e}n, Philipp Koch, Ngoc~QK Duong, Ian
  McLoughlin, and Alfred Mertins.
\newblock Self-attention generative adversarial network for speech enhancement.
\newblock In {\em ICASSP 2021-2021 IEEE International Conference on Acoustics,
  Speech and Signal Processing (ICASSP)}, pages 7103--7107. IEEE, 2021.

\bibitem{zhao2020exploring}
Hengshuang Zhao, Jiaya Jia, and Vladlen Koltun.
\newblock Exploring self-attention for image recognition.
\newblock In {\em Proceedings of the IEEE/CVF Conference on Computer Vision and
  Pattern Recognition}, pages 10076--10085, 2020.

\bibitem{amcg_2015}
Imperial College~London AMCG.
\newblock Fluidity manual v4.1.12, Apr 2015.

\bibitem{xiao2015nonrbf}
D~Xiao, F~Fang, C~Pain, and G~Hu.
\newblock Non-intrusive reduced-order modelling of the navier--stokes equations
  based on {RBF} interpolation.
\newblock {\em International Journal for Numerical Methods in Fluids},
  79(11):580--595, 2015.

\bibitem{pain2005three}
CC~Pain, MD~Piggott, AJH Goddard, F~Fang, GJ~Gorman, DP~Marshall, MD~Eaton,
  PW~Power, and CRE De~Oliveira.
\newblock Three-dimensional unstructured mesh ocean modelling.
\newblock {\em Ocean Modelling}, 10(1-2):5--33, 2005.

\bibitem{shin2004gravity}
JO~Shin, SB~Dalziel, and PF~Linden.
\newblock Gravity currents produced by lock exchange.
\newblock {\em Journal of Fluid Mechanics}, 521:1--34, 2004.

\end{thebibliography}

\end{document}